\documentclass[aps,prd,twocolumn,superscriptaddress]{revtex4}
\usepackage{epsfig,epsf}
\usepackage{amsmath}
\usepackage{amsthm}
\usepackage{amsfonts}
\usepackage{amssymb}
\usepackage{dsfont}
\usepackage{multirow}
\usepackage{appendix}
\usepackage{slashed}
\usepackage[active]{srcltx}
\usepackage{psfrag}

\begin{document}

\title{Treating $Z_c(3900)$ and $Z(4430)$ as the ground-state and first
radially excited tetraquarks}
\date{\today}
\author{S.~S.~Agaev}
\affiliation{Institute for Physical Problems, Baku State University, Az--1148 Baku,
Azerbaijan}
\author{K.~Azizi}
\affiliation{Department of Physics, Do\v{g}u\c{s} University, Acibadem-Kadik\"{o}y, 34722
Istanbul, Turkey}
\author{H.~Sundu}
\affiliation{Department of Physics, Kocaeli University, 41380 Izmit, Turkey}

\begin{abstract}
Exploration of the resonances $Z_c(3900)$ and $Z(4430)$ are performed by
assuming that they are ground-state and first radial excitation of the same
tetraquark with $J^{P}=1^{+}$. The mass and current coupling of the $%
Z_c(3900)$ and $Z(4430)$ states are calculated using QCD two-point sum rule
method by taking into account vacuum condensates up to eight dimensions. We
investigate the vertices $Z_cM_hM_l$ and $ZM_hM_l$, with $M_h$ and $M_l $
being the heavy and light mesons,  and evaluate the strong couplings $%
g_{Z_cM_hM_l}$ and $g_{ZM_hM_l}$ using QCD sum rule on the light cone. The
extracted couplings allow us to find the partial width of the decays $%
Z_c(3900) \to J/\psi \pi; \,\psi^{\prime} \pi;\, \eta_c \rho$ and $Z(4430)
\to \psi^{\prime} \pi;\, J/\psi \pi;\, \eta_c^{\prime}\rho;\, \eta_c\rho$,
which may help in comprehensive investigation of these resonances. We
compare width of the decays of $Z_c(3900)$ and $Z(4430)$ resonances with
available experimental data as well as  existing theoretical
predictions.
\end{abstract}

\maketitle

\section{Introduction}

The discoveries of the charged $Z(4430)$ and $Z_c(3900)$ resonances had
important consequences for the physics of multi-quark hadrons, because they
could not be interpreted as neutral $\bar c c$ charmonia and became real
candidates to tetraquark states. The $Z^{\pm}(4430)$ states were observed by
the Belle Collaboration in $B$ meson decays $B \to K\psi^{\prime} \pi^{\pm}$
as  resonances in the $\psi^{\prime}\pi^{\pm}$ invariant mass distribution
\cite{Choi:2007wga}. The resonances $Z^{+}(4430)$ and $Z^{-}(4430)$ were
detected and studied later again by Belle in the processes $B \to
K\psi^{\prime} \pi^{+}$ \cite{Mizuk:2009da} and $B^{0} \to
K^{+}\psi^{\prime} \pi^{-}$ \cite{Chilikin:2013tch}, respectively. An
evidence for $Z(4430)$ resonance decaying to $J/\psi \pi$ was found by the
same collaboration in the process $\bar{B}^0 \to J/\psi K^{-}\pi^{+}$ \cite%
{Chilikin:2014bkk}. The available experimental information allowed the Belle
Collaboration, apart from the masses and decay widths of these resonances, to
fix also their spin-parity $J^{P}=1^{+}$ as a most favorable assumption
among the $0^{-},\, 1^{\pm}$ and $2^{\pm}$ options. The parameters of $%
Z^{-}(4430)$ were measured in the $B^{0} \to K^{+}\psi^{\prime} \pi^{-}$
decay by the LHCb Collaboration with the results
\begin{equation}
M=(4475\pm 7_{-25}^{+15})\, \mathrm{MeV},\, \Gamma=(172\pm 13_{-34}^{+37})\,
\mathrm{MeV},  \label{eq:LHCdata}
\end{equation}
where its spin-parity was unambiguously determined to be $1^{+}$ \cite%
{Aaij:2014jqa,Aaij:2015zxa}.

Other members of the charged tetraquarks family, namely $Z_c^{\pm}(3900)$
were discovered by the BESIII Collaboration in the process $e^{+}e^{-} \to
J/\psi \pi^{+} \pi^{-}$, as resonances in the $J/\psi \pi^{\pm}$ invariant
mass distributions with the parameters
\begin{equation}
M=(3899.0\pm 3.6 \pm 4.9)\, \mathrm{MeV},\, \Gamma=(46\pm 10 \pm 20)\,
\mathrm{MeV},
\end{equation}
and spin-parity $J^{P}=1^{+}$ \cite{Ablikim:2013mio}. These structures were
observed also by the Belle and CLEO collaborations, as well (see, Refs.\
\cite{Liu:2013dau,Xiao:2013iha}). Recently, BESIII announced the observation
of the neutral $Z_c^{0}(3900)$ state in the process $e^{+}e^{-}\rightarrow
\pi ^{0}Z_{c}^{0}\rightarrow \pi ^{0}\pi^{0}J/\psi $ \cite{Ablikim:2015tbp}.

Theoretical investigations of the $Z(4430)$ and $Z_c(3900)$ resonances
embrace a variety of models and computational schemes \cite%
{Chen:2016qju,Esposito:2016noz}. The aim is to reveal their internal
quark-gluon structure and determine their parameters, such as the masses, current
couplings (pole residues) and width of decay modes. Thus, $Z(4430)$ was
interpreted as the diquark-antidiquark \cite%
{Liu:2008qx,Ebert:2008kb,Bracco:2008jj,Maiani:2008zz,Wang:2010rt,Maiani:2014,Wang:2014vha},
 molecular state \cite%
{Lee:2007gs,Liu:2008xz,Braaten:2007xw,Branz:2010sh,Goerke:2016hxf}, the
threshold effect \cite{Rosner:2007mu} and hadro-charmonium composite \cite%
{Dubynskiy:2008mq}.

The situation formed around the theoretical interpretation of the $Z_c(3900)$
resonance does not differ considerably from activities intending to explain
features of $Z(4430)$. Indeed, there are attempts to treat it as the tightly
bound diquark-antidiquark state \cite%
{Dias:2013xfa,Wang:2013vex,Deng:2014gqa,Agaev:2016dev}, as the four-quark
bound state composed of conventional mesons \cite%
{Wang:2013daa,Wilbring:2013cha,Dong:2013iqa,Ke:2013gia,Gutsche:2014zda,Esposito:2014hsa, Chen:2015igx,Gong:2016hlt,Ke:2016owt}%
, or as the threshold cusp \cite{Swanson:2014tra,Ikeda:2016zwx}.

The interesting idea was suggested in Ref.\ \cite{Maiani:2014} to consider
the $Z_c(3900)$ and $Z(4430)$ resonances as the ground and first radially
excited states of the same diquark-antidiquark multiplet. This assumption
was motivated by the main decay channels of these resonances,
\begin{equation}
Z_c^{\pm}(3900) \to J/\psi \pi^{\pm},\,\, Z^{\pm}(4430) \to \psi^{\prime}
\pi^{\pm},
\end{equation}
and also by observation that the mass difference between the $1S$ and $2S$
states $m_{\psi^{\prime}}-m_{J/\psi}$ is approximately equal to the mass
splitting $m_{Z}-m_{Z_c}$. This idea was realized within the
diquark-antidiquark model, and in the context of QCD sum rule approach in
Ref.\ \cite{Wang:2014vha}, where the masses and pole residues of $Z_c(3900)$
and $Z(4430)$ were obtained. The performed analysis in this work  seems
to confirm a suggestion made there. It should be noted that the decay modes of
the resonances $Z_c(3900)$ and $Z(4430)$, which contain an important
dynamical information on the structure of these states, were not considered
within this scheme.

The mass and decay constant (current coupling) are important spectroscopic
parameters of a conventional hadron or an exotic multi-quark state, which
should be measured and calculated first of all. Therefore, not surprisingly
all theoretical models and schemes proposed to explain the internal
structure of tetraquarks and their properties start from analysis and
computation of these parameters. Only after obtaining reasonable
predictions for the mass and current coupling a model may claim to be a
correct theory of a tetraquark candidate. But this is not enough to make
robust conclusions on the nature of observed resonances. Indeed,
experimental investigations include measurements of both the masses and widths
of the observed resonances, and provide additional information on their
spins and parities.

Almost all models of the resonances $Z_c(3900)$ and $Z(4430)$ correctly
predict their masses. In some of theoretical papers the decay channels of
these states were addressed, as well. Thus, the decays of the $Z^{\pm}(4430)$
states $Z^{\pm}(4430) \to J/\psi \pi^{\pm};\, \psi^{\prime} \pi^{\pm}$ were
investigated within a phenomenological Lagrangian approach by interpreting
it as a molecular state with the structure $D_1(2420)\bar D^{\star} + h.c.$
in Ref.\ \cite{Branz:2010sh}. Unfortunately, in this paper $Z(4430)$ was
treated as a state with spin-parity $J^{P}=0^{-},\,1^{-}$ excluded by recent
measurements. The same decay modes $Z^{+}(4430) \to J/\psi \pi^{+};\,
\psi^{\prime} \pi^{+}$ were revisited in a covariant quark model in Ref.\
\cite{Goerke:2016hxf}.

The different decay modes of the $Z_c(3900)$ state in a diquark-antidiquark
model were analyzed in Refs.\ \cite{Dias:2013xfa} and \cite{Agaev:2016dev}.
The $Z_c^{+}(3900)\to J/\psi \pi^{+};\,\eta_c \rho;\, D^{+}\bar {D}^{\star
0} $ decays' widths were computed in Ref.\ \cite{Dias:2013xfa} using the
three-point sum rule method, whereas in Ref.\ \cite{Agaev:2016dev} widths of
the $Z_c^{+}(3900)\to J/\psi \pi^{+};\,\eta_c \rho$ decays were found by
means of the light cone sum rule (LCSR) approach and a technique of the
soft-meson approximation. In both of these works $Z_c(3900)$ was considered
as a state with the spin-parities $J^{PC}=1^{+-}$.

The decays of the $Z_c^{\pm}(3900)$ resonances were a subject of studies in
the framework of alternative methods \cite%
{Dong:2013iqa,Gutsche:2014zda,Goerke:2016hxf}, as well. Thus, the decay channels
$Z_c(3900) \to J/\psi \pi;\, \psi^{\prime} \pi; h_c(1P) \pi $ were
calculated in a phenomenological Lagrangian approach by modeling $Z_c(3900)$
as a hadronic molecule $\bar DD^{\star}$ with $J^{P}=1^{+}$ \cite%
{Dong:2013iqa}. The radiative and leptonic decays $Z_c^{+}(3900) \to J/\psi
\pi^{+}\gamma$ and $J/\psi \pi^{+}l^{+}l^{-},\, l=(e,\,\mu)$ in the context
of the same method were considered in Ref.\ \cite{Gutsche:2014zda}. The
widths of the decay modes $Z_c^{+}(3900) \to J/\psi \pi^{+};\, \eta_c
\rho^{+}; \bar{D}^{0}D^{\star +};\, \bar{D}^{\star 0}D^{+}$ were extracted
in Ref.\ \cite{Goerke:2016hxf} in a covariant quark model. Let us note also
the work \cite{Ke:2016owt}, where the $Z_c(3900) \to h_c\pi $ decay was
analyzed in the light front model.

In the present study we are going to calculate the masses and current
couplings of the $Z_c(3900)$ and $Z(4430)$ resonances, and investigate some
of their decay modes. We assume that these resonances are the ground and first
radially excited states of the tetraquark with $J^{P}=1^{+}$, i.e. we
consider them as the axial-vector members of the $1S$ and $2S$ tetraquark
multiplets. We will evaluate the mass and current coupling of the excited $%
Z(4430)$ state and width of the process $Z(4430) \to \psi^{\prime} \pi$,
which is the main decay channel of $Z(4430)$ to examine correctness of the
suggestion made on its nature. Other decay modes of the $Z(4430)$ resonance,
namely $Z(4430) \to J/\psi \pi;\, \eta_c^{\prime}\rho;\, \eta_c\rho$ will be
analyzed, as well. We  will also calculate the $Z_c(3900)$ resonance's
parameters and its decay widths with higher accuracy than it was done in our
previous work \cite{Agaev:2016dev}. We will include into analysis also the
decay mode $Z_c(3900) \to \psi^{\prime} \pi$, which was not considered in
the previous paper.

This work is organized in the following form. In Sec.\ \ref{sec:Mass} we
derive the spectral density $\rho^{\mathrm{QCD}}(s)$ from the two-point QCD
sum rule by including condensates up to eight dimensions. This allows us to
evaluate the mass and current coupling of the $Z_c(3900)$ and $Z(4430)$ with
desired accuracy. The Sec.\ \ref{sec:StrongVer1} is devoted to decays of the
$Z_c(3900)$ and $Z(4430)$ states. Here we calculate relevant spectral
densities with dimension-eight accuracy and find the width of the decays
under consideration. Appendix contains the quark propagators used in this
work, as well as explicit expressions of the spectral densities used in
computation of the strong couplings.

%%%%%%%%%%%%%%%%%%%%%%%%%%%%%%%%%%%%%%%%%%%%%%%%%%%%%%%%%%%%%%%%%%%%%%%%%%%%

\section{Masses and current couplings of the $Z_c(3900)$ and $Z(4430)$
resonances}

\label{sec:Mass}
%%%%%%%%%%%%%%%%%%%%%%%%%%%%%%%%%%%%%%%%%%%%%%%%%%%%%%%%%%%%%%%%%%%%%%%%%%%%
In this section we calculate the masses and current couplings of the resonances
$Z_c(3900)$ and $Z(4430)$. We consider the positively charged states with
the quark content $\bar{c}cu\bar{d}$, but due to the exact chiral limit
adopted in the present work, the parameters of the resonances with opposite
charges do not differ from each other.

The masses and current couplings of the resonances under consideration can be extracted from
analysis of the correlation function
\begin{equation}
\Pi _{\mu \nu }(p)=i\int d^{4}xe^{ip\cdot x}\langle 0|\mathcal{T}\{J_{\mu
}^{Z}(x)J_{\nu }^{Z\dagger }(0)\}|0\rangle ,  \label{eq:CorrF1}
\end{equation}%
where the interpolating current with the quantum numbers $J^{PC}=1^{+-}$ is
given by the following expression
\begin{eqnarray}
J_{\nu }^{Z}(x) &=&\frac{i\epsilon \tilde{\epsilon}}{\sqrt{2}}\left\{ \left[
u_{a}^{T}(x)C\gamma _{5}c_{b}(x)\right] \left[ \overline{d}_{d}(x)\gamma
_{\nu }C\overline{c}_{e}^{T}(x)\right] \right.  \notag \\
&&\left. -\left[ u_{a}^{T}(x)C\gamma _{\nu }c_{b}(x)\right] \left[ \overline{%
d}_{d}(x)\gamma _{5}C\overline{c}_{e}^{T}(x)\right] \right\} .
\label{eq:Curr}
\end{eqnarray}%
Here we have introduced the notations $\epsilon =\epsilon _{abc}$ and $%
\tilde{\epsilon}=\epsilon _{dec}$. In Eq.\ (\ref{eq:Curr}) $a,b,c,d,e$ are
color indices and $C$ is the charge conjugation operator.

We first derive the sum rules for the mass $m_{Z_c}$ and current coupling $%
f_{Z_c}$ of the ground state tetraquark $Z_c$. To this end, we use the
"ground-state + continuum" approximation by including the $Z(4430)$ state
into the list of "higher resonances" and extract corresponding sum rules.
The mass and current coupling of $Z_c$ evaluated from these expressions are
considered as input parameters in the sum rules for the excited $Z(4430)$
tetraquark.

At the next stage of calculations we adopt the "ground-state+radially
excited state+continuum" scheme, and perform the required standard
manipulations: we derive the phenomenological side of the sum rules by
inserting into the correlation function  full sets of relevant states, by
isolating contributions of the $Z_c$ and $Z$ resonances and carrying out the
integration over $x$. As a result, for $\Pi _{\mu \nu }^{\mathrm{Phys}}(p)$
we get
\begin{eqnarray}
&&\Pi _{\mu \nu }^{\mathrm{Phys}}(p)=\frac{\langle 0|J_{\mu
}^{Z}|Z_{c}(p)\rangle \langle Z_{c}(p)|J_{\nu }^{Z\dagger }|0\rangle }{%
m_{Z_{c}}^{2}-p^{2}}  \notag \\
&&+ \frac{\langle 0|J_{\mu }^{Z}|Z(p)\rangle \langle Z(p)|J_{\nu }^{Z\dagger
}|0\rangle }{m_{Z}^{2}-p^{2}}+\ldots  \label{Phys1}
\end{eqnarray}%
where $m_{Z_c}$ and $m_{Z}$ are the masses of  $Z_c(3900)$ and $Z(4430)$ states,
respectively. The dots in Eq.\ (\ref{Phys1}) denote contributions arising from
higher resonances and continuum states.

In order to finish computations of the sum rules' phenomenological side we
introduce the current couplings $f_{Z_{c}}$ and $f_{Z}$ through the matrix
elements
\begin{equation}
\langle 0|J_{\mu }^{Z}|Z_{c}\rangle =f_{Z_{c}}m_{Z_{c}}\varepsilon _{\mu
},\,\,\langle 0|J_{\mu }^{Z}|Z\rangle =f_{Z}m_{Z}\tilde{\varepsilon}_{\mu },
\label{eq:Res}
\end{equation}%
with $\varepsilon _{\mu }$ and $\tilde{\varepsilon}_{\mu }$ being the
polarization vectors of the $Z_{c}$ and $Z$ states, respectively. Then the
function $\Pi _{\mu \nu }^{\mathrm{Phys}}(p)$ can be written as
\begin{eqnarray}
&&\Pi _{\mu \nu }^{\mathrm{Phys}}(p)=\frac{m_{Z_{c}}^{2}f_{Z_{c}}^{2}}{%
m_{Z_{c}}^{2}-p^{2}}\left( -g_{\mu \nu }+\frac{p_{\mu }p_{\nu }}{%
m_{Z_{c}}^{2}}\right)  \notag \\
&&+\frac{m_{Z}^{2}f_{Z}^{2}}{m_{Z}^{2}-p^{2}}\left( -g_{\mu \nu }+\frac{%
p_{\mu }p_{\nu }}{m_{Z}^{2}}\right) +\ldots  \label{eq:CorM}
\end{eqnarray}%
The Borel transformation applied to Eq.\ (\ref{eq:CorM}) yields%
\begin{eqnarray}
&&\mathcal{B}_{p^{2}}\Pi _{\mu \nu }^{\mathrm{Phys}%
}(p)=m_{Z_{c}}^{2}f_{Z_{c}}^{2}e^{-m_{Z_{c}}^{2}/M^{2}}\left( -g_{\mu \nu }+%
\frac{p_{\mu }p_{\nu }}{m_{Z_{c}}^{2}}\right)  \notag \\
&&+m_{Z}^{2}f_{Z}^{2}e^{-m_{Z}^{2}/M^{2}}\left( -g_{\mu \nu }+\frac{p_{\mu
}p_{\nu }}{m_{Z}^{2}}\right) +\ldots ,
\end{eqnarray}%
where $M^{2}$ is the Borel parameter.

The QCD side of the sum rules $\Pi _{\mu \nu }^{\mathrm{QCD}}(q)$ can be
determined using the interpolating current $J_{\nu}^{Z}$ and quark
propagators, explicit expressions of which can be found in Appendix. Thus,
after contracting the heavy and light quark fields in Eq.\ (\ref{eq:CorrF1})
we get
\begin{eqnarray}
&&\Pi _{\mu \nu }^{\mathrm{QCD}}(p)=-\frac{i}{2}\int d^{4}xe^{ipx}\epsilon
\tilde{\epsilon}\epsilon ^{\prime }\tilde{\epsilon}^{\prime }\left\{ \mathrm{%
Tr}\left[ \gamma _{5}\widetilde{S}_{u}^{aa^{\prime }}(x)\right. \right.
\notag \\
&&\times \left. \gamma _{5}S_{c}^{bb^{\prime }}(x)\right] \mathrm{Tr}\left[
\gamma _{\mu }\widetilde{S}_{c}^{e^{\prime }e}(-x)\gamma _{\nu
}S_{d}^{d^{\prime }d}(-x)\right]   \notag \\
&&-\mathrm{Tr}\left[ \gamma _{\mu }\widetilde{S}_{c}^{e^{\prime
}e}(-x)\gamma _{5}S_{d}^{d^{\prime }d}(-x)\right] \mathrm{Tr}\left[ \gamma
_{\nu }\widetilde{S}_{u}^{aa^{\prime }}(x)\gamma _{5}S_{c}^{bb^{\prime }}(x)%
\right]   \notag \\
&&-\mathrm{Tr}\left[ \gamma _{5}\widetilde{S}_{u}^{a^{\prime }a}(x)\gamma
_{\mu }S_{c}^{b^{\prime }b}(x)\right] \mathrm{Tr}\left[ \gamma _{5}%
\widetilde{S}_{c}^{e^{\prime }e}(-x)\gamma _{\nu }S_{d}^{d^{\prime }d}(-x)%
\right]   \notag \\
&&\left. +\mathrm{Tr}\left[ \gamma _{\nu }\widetilde{S}_{u}^{aa^{\prime
}}(x)\gamma _{\mu }S_{c}^{bb^{\prime }}(x)\right] \mathrm{Tr}\left[ \gamma
_{5}\widetilde{S}_{c}^{e^{\prime }e}(-x)\gamma _{5}S_{d}^{d^{\prime }d}(-x)%
\right] \right\}, \notag \\
{}
 \label{eq:CorrF2}
\end{eqnarray}%
where%
\begin{equation*}
\widetilde{S}_{c(q)}^{ij}(x)=CS_{c(q)}^{ij\mathrm{T}}(x)C.
\end{equation*}

The function $\Pi _{\mu \nu }^{\mathrm{QCD}}(p)$ has the following
decomposition over the Lorentz structures%
\begin{equation}
\Pi _{\mu \nu }^{\mathrm{QCD}}(p)=\Pi ^{\mathrm{QCD}}(p^{2})(-g_{\mu \nu })+%
\widetilde{\Pi }^{\mathrm{QCD}}(p^{2})p_{\mu }p_{\nu }.
\end{equation}%
The required QCD sum rules for the parameters of $Z(4430)$ can be obtained
after equating the same structures in both $\Pi _{\mu \nu }^{\mathrm{Phys}%
}(p)$ and $\Pi _{\mu \nu }^{\mathrm{QCD}}(p)$. For our purposes the
convenient structures are terms $\sim (-g_{\mu \nu })$, which we employ in
further calculations.

The invariant amplitude corresponding to the structure $-g_{\mu \nu }$ in $%
\Pi _{\mu \nu }^{\mathrm{Phys}}(p)$ has the simple form. The similar
function $\Pi ^{\mathrm{QCD}}(p^{2})$ can be represented as the dispersion
integral
\begin{equation}
\Pi^{\mathrm{QCD}}(p^{2})=\int_{4m_{c}^{2}}^{\infty }\frac{\rho ^{\mathrm{QCD%
}}(s)}{s-p^{2}}ds,
\end{equation}%
where $\rho^{\mathrm{QCD}}(s)$ is the two-point spectral density. Methods to
derive $\rho^{\mathrm{QCD}}(s)$ as the imaginary part of the correlation
function are well known, therefore we skip here details of standard
calculations, and also refrain from providing its explicit expression.

We apply the Borel transformation on the variable $p^{2}$ to both the
phenomenological and QCD sides of the equality and subtract the contributions
of the higher resonances and continuum states by invoking the assumption on
the quark-hadron duality. After simple manipulations we derive the sum rules
for the mass and current coupling of the excited $Z(4430)$ state:
\begin{equation}
m_{Z}^{2}=\frac{\int_{4m_{c}^{2}}^{s_{0}^{\ast }}\rho ^{\mathrm{QCD}%
}(s)se^{-s/M^{2}}ds-f_{Z_{c}}^{2}m_{Z_{c}}^{4}e^{-m_{Z_{c}}^{2}/M^{2}}}{%
\int_{4m_{c}^{2}}^{s_{0}^{\ast }}\rho ^{\mathrm{QCD}%
}(s)e^{-s/M^{2}}ds-f_{Z_{c}}^{2}m_{Z_{c}}^{2}e^{-m_{Z_{c}}^{2}/M^{2}}},
\label{eq:MassEx}
\end{equation}%
and%
\begin{eqnarray}
f_{Z}^{2}&=&\frac{1}{m_{Z}^{2}}\left[ \int_{4m_{c}^{2}}^{s_{0}^{\ast }}\rho
^{\mathrm{QCD}}(s)e^{(m_{Z}^{2}-s)/M^{2}}ds \right.  \notag \\
&&\left.-f_{Z_{c}}^{2}m_{Z_{c}}^{2}e^{(m_{Z}^{2}-m_{Z_{c}}^{2})/M^{2}}\right]%
,  \label{eq:CoupEx}
\end{eqnarray}%
where $s_{0}^{\ast }$ is the continuum threshold parameter, which separates
the contributions of the "$Z_{c}$ +$Z$" states from the contributions due to
"higher resonances and continuum". As we have emphasized above the mass and
current coupling of $Z_{c}$ enter into Eqs.\ (\ref{eq:MassEx}) and (\ref%
{eq:CoupEx}) as the input parameters. The mass of the $Z_c$ state can be
found from the sum rule
\begin{equation}
m_{Z_{c}}^{2}=\frac{\int_{4m_{c}^{2}}^{s_{0}}ds\rho ^{\mathrm{QCD}%
}(s)se^{-s/M^{2}}}{\int_{4m_{c}^{2}}^{s_{0}}ds\rho^{%
\mathrm{QCD}} (s)e^{-s/M^{2}}},
\label{eq:MassGS}
\end{equation}%
whereas to extract the numerical value of the decay constant $f_{Z_{c}}$ we
employ the formula
\begin{equation}
f_{Z_{c}}^{2}=\frac{1}{m_{Z_{c}}^{2}}\int_{4m_{c}^{2}}^{s_{0}}ds\rho ^{%
\mathrm{QCD}}(s)e^{(m_{Z_{c}}^{2}-s)/M^{2}}.  \label{eq:CoupGS}
\end{equation}
In Eqs.\ (\ref{eq:MassGS}) and (\ref{eq:CoupGS}) $s_0$ is the continuum
threshold, which divides the contributions of the ground-state $Z_c(3900)$
and higher resonances and continuum. It is evident, that sum rules depend on
the same spectral density $\rho ^{\mathrm{QCD}}(s)$, and the continuum
threshold has to obey $s_0 < s_0^{\star}$.

The expressions obtained in the present work contain the vacuum expectations
values of the different operators, which are input parameters in the
numerical calculations. These vacuum condensates are well known: for the
quark and mixed condensates we use $\langle \bar{q}q \rangle =-(0.24\pm
0.01)^3~\mathrm{GeV}^3$ and $\langle \overline{q}g_{s}\sigma
Gq\rangle=m_{0}^2\langle \bar{q}q\rangle$, where $m_{0}^2 =(0.8\pm0.1)~%
\mathrm{GeV}^2$, whereas for the gluon condensates we utilize $%
\langle\alpha_sG^2/\pi\rangle=(0.012\pm0.004)~\mathrm{GeV}^4 $ and $\langle
g_{s}^3G^3\rangle =(0.57\pm0.29)~\mathrm{GeV}^6 $.

The sum rules depend also on the auxiliary parameters $M^2$ and $%
s_0(s_0^{\star})$, which have to meet requirements of the sum rule
computations. In other words, the bounds of working region for the Borel
parameter are fixed by the convergence of the operator product expansion and
dominance of the pole contribution to the whole expression. Additionally,
regions within of which the parameters $M^2$ and $s_0$ can be varied should
provide stability domains of extracted quantities on the parameters $M^2$
and $s_0$. Performed analysis allows us to fix regions of the parameters $%
M^2 $ and $s_0$, where the aforementioned conditions are satisfied.
Numerical results of our calculations are collected in Table \ref{tab:Results1},
where we write down
not only the mass and current couplings of the resonances $Z(4430)$ and $%
Z_c(3900)$, but also regions of the parameters used in their evaluation. It
is seen that $m_{Z_c}$ is in excellent agreement with the experimental
data. It also almost coincide with our previous result for $m_{Z_c}$
obtained in Ref.\ \cite{Agaev:2016dev}. Prediction for $m_Z$ is smaller than
the corresponding LHCb result, but within errors of calculations it is
compatible with measurements.

In Figs. \ref{fig:Mass} and \ref{fig:Coup} we demonstrate the dependence of $%
m_Z$ and $f_Z$ on $M^2$ at fixed $s_0$, and as functions of $s_0$ for chosen
values of $M^2$. As is seen, the mass of the $Z(4430)$ resonance is rather
stable against variations both of $M^2$ and $s_0$, whereas a sensitivity of $%
f_Z$ to changes of the auxiliary parameters is higher. The explanation here
is quite simple: indeed, the sum rules for the mass of the resonances are
given as ratios of integrals over the spectral densities $\rho(s)$ and $%
s\rho(s)$, which considerably reduce effects due to variations of $M^2$ and $%
s_0$. Contrary, the current couplings depend on mentioned integrals
themselves, and therefore undergone relatively sizable changes. Thus,
theoretical errors for $f_{Z_c}$ and $f_Z$ stemming from uncertainties of $%
M^2$ and $s_0$, and other input parameters equal to $\sim 19 \%$ and $\sim
28 \% $ of the central values, respectively. Nevertheless, they remain
within allowed ranges for theoretical errors inherent to sum rule
computations which may amount up to $30 \%$ of predictions.
\begin{table}[tbp]
\begin{tabular}{|c|c|c|}
\hline\hline
Resonance & $Z_c(3900)$ & $Z(4430)$ \\ \hline\hline
$M^2 ~(\mathrm{GeV}^2$) & $3-6$ & $3-6$ \\ \hline
$s_0(s_0^{\star}) ~(\mathrm{GeV}^2$) & $4.2^2-4.4^2$ & $4.8^2-5.2^2$ \\
\hline
$m_{Z} ~(\mathrm{MeV})$ & $3901^{+125}_{-148} $ & $4452^{+182}_{-228} $ \\
\hline
$f_{Z}\cdot 10^{2} ~(\mathrm{GeV}^4)$ & $0.42^{+0.07}_{-0.09}$ & $%
1.48^{+0.31}_{-0.42}$ \\ \hline\hline
\end{tabular}%
\caption{The masses and current couplings of the $Z_c(3900)$ and $Z(4430) $
resonances evaluated in this work.}
\label{tab:Results1}
\end{table}

\begin{widetext}

\begin{figure}[h!]
\begin{center}
\includegraphics[totalheight=6cm,width=8cm]{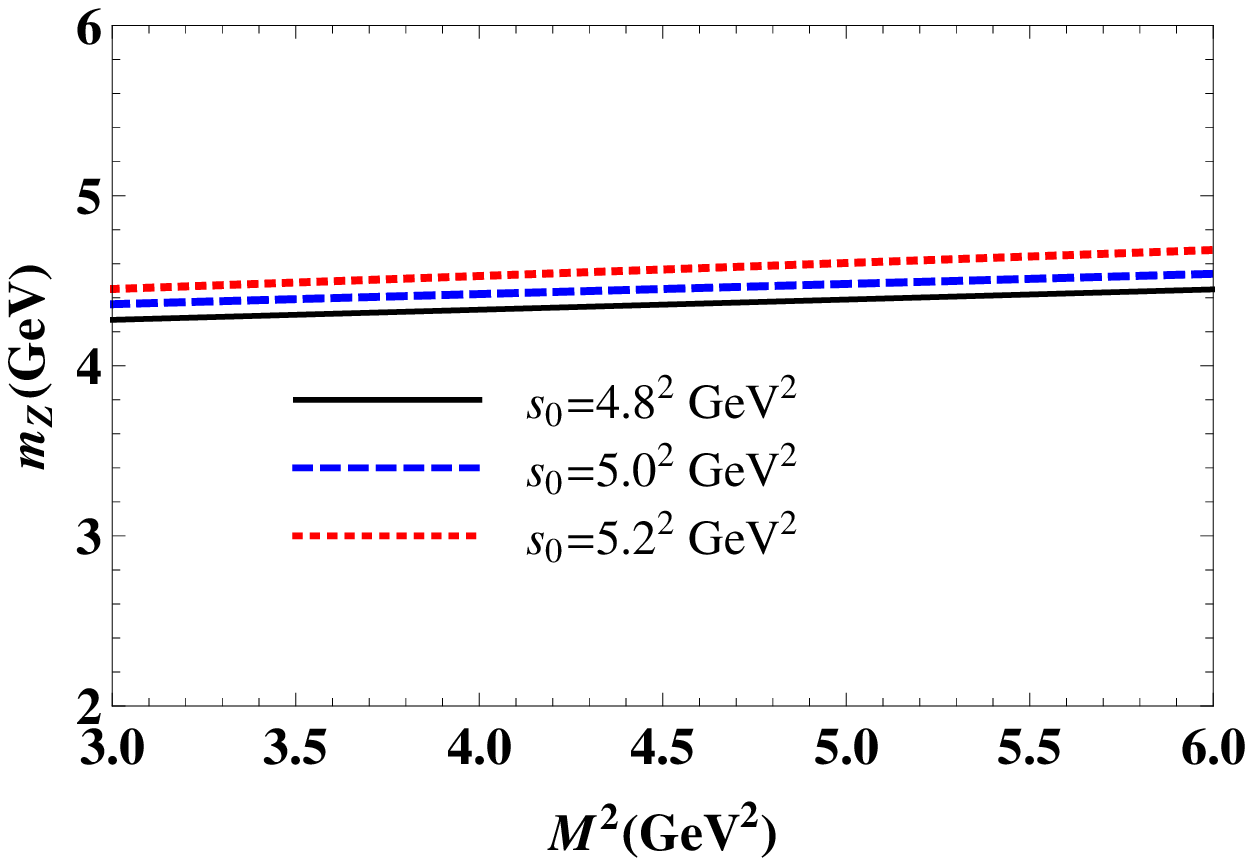}\,\, %
\includegraphics[totalheight=6cm,width=8cm]{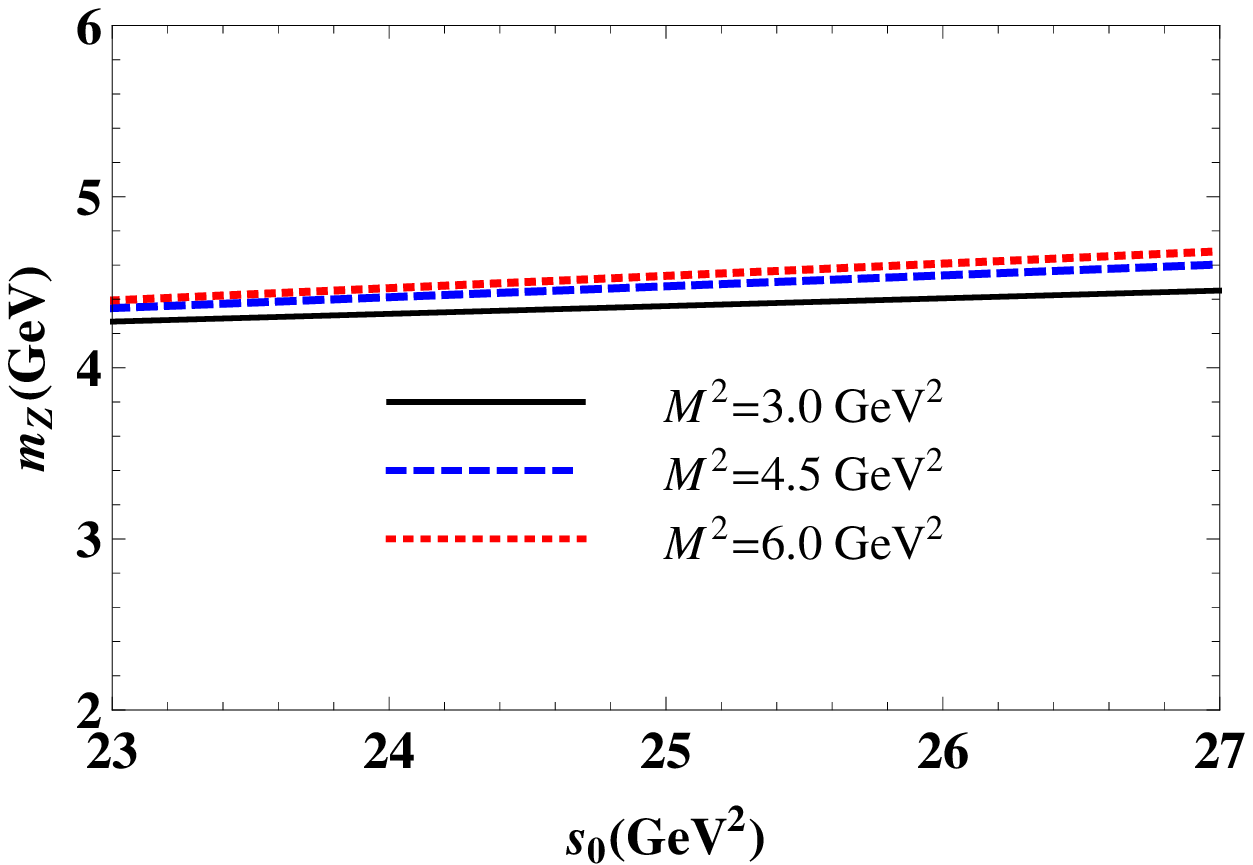}
\end{center}
\caption{ The mass of the state $Z(4430)$ as a function of the Borel parameter
$M^2$ at fixed $s_0$ (left panel), and as a function of the continuum threshold
$s_0$ at fixed $M^2$ (right panel).}
\label{fig:Mass}
\end{figure}
\begin{figure}[h!]
\begin{center}
\includegraphics[totalheight=6cm,width=8cm]{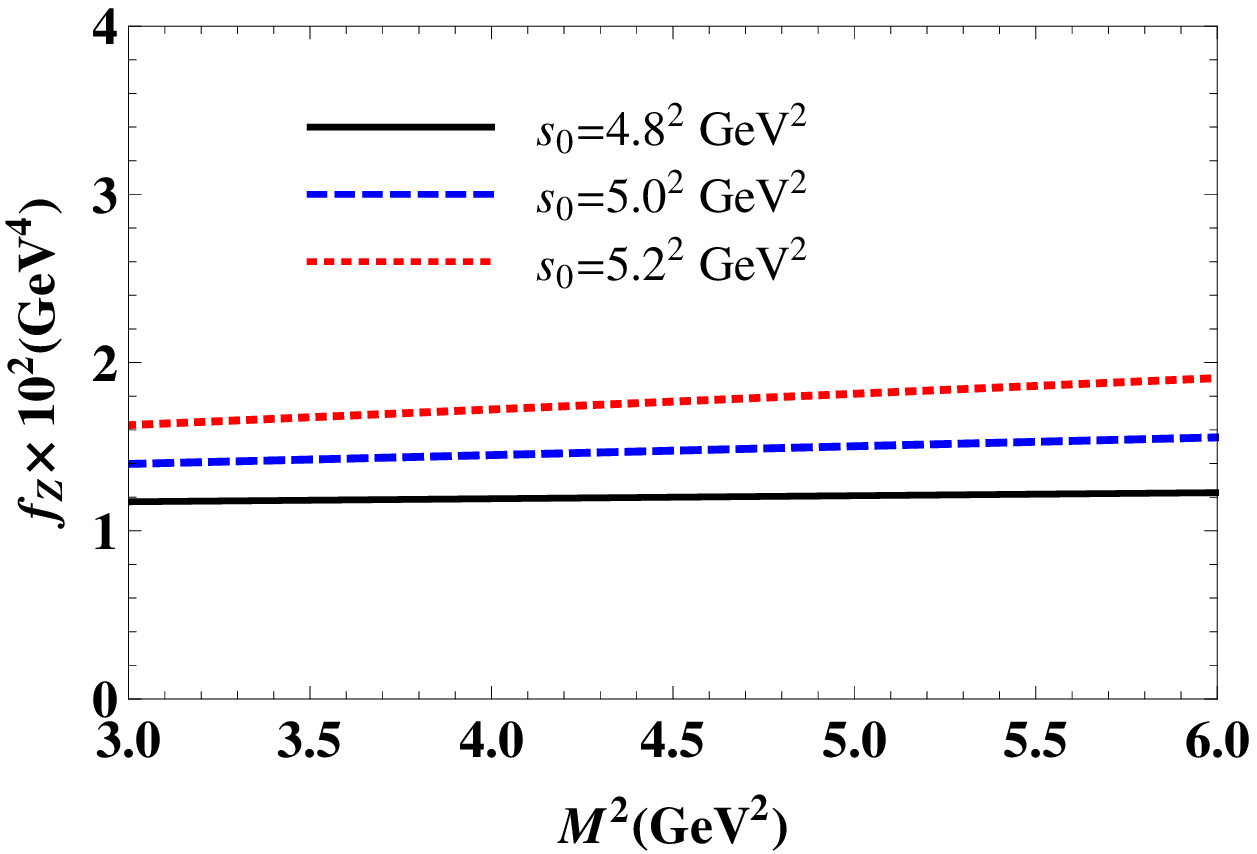}\,\, %
\includegraphics[totalheight=6cm,width=8cm]{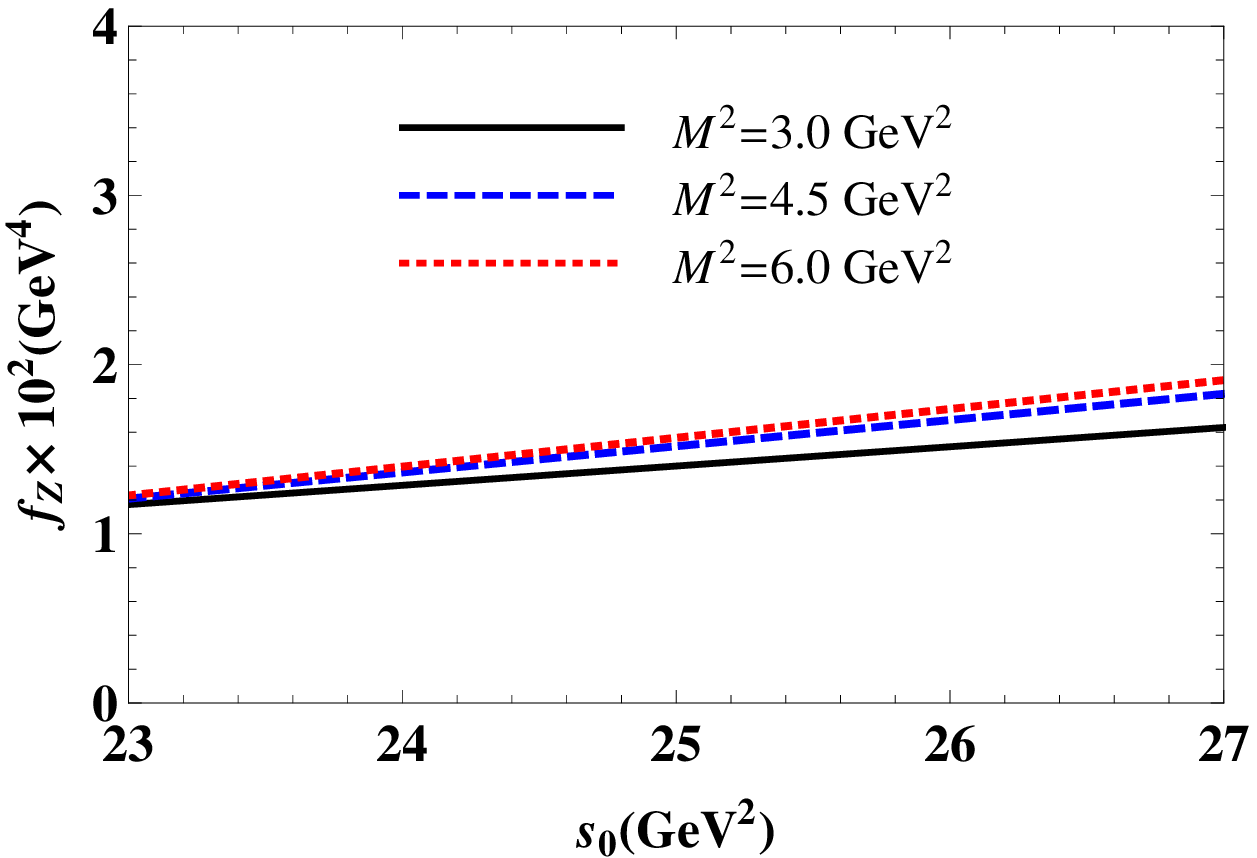}
\end{center}
\caption{ The dependence of the current coupling $f_Z$ of the $Z(4430)$ resonance
on the Borel parameter at chosen values of $s_0$ (left panel), and on  $s_0$ at
fixed $M^2$  (right panel).}
\label{fig:Coup}
\end{figure}

\end{widetext}

\section{Decays of the resonances $Z_c(3900)$ and $Z(4430)$}

\label{sec:StrongVer1} The masses of $Z_c(3900)$ and $Z(4430)$ extracted in
the previous section can be used to fix kinematically allowed decay modes of
these resonances. Moreover, their masses and current couplings enter as
input parameters to expressions of the strong couplings describing vertices $%
Z_cM_hM_l$ and $ZM_hM_l$, and appear also in formulas for decay widths.

The resonances $Z_c(3900)$ and $Z(4430)$ may decay through different
channels. In this work we restrict ourselves by analysis of their decays to $%
J/\psi \pi$, $\psi^{\prime} \pi$, and $\eta_c \rho $, $\eta_c^{\prime} \rho$
mesons: from Table \ref{tab:Param}, where we provide masses and decay
constants of the relevant mesons, it is easy to realize that these decays
are among kinematically allowed channels.

In this work we treat the $Z(4430)$ tetraquark as the first radial
excitation of $Z_c(3900)$. Additionally, $\psi^{\prime}$ and $%
\eta_c^{\prime} $ are the first radial excitations of the mesons $J/\psi$
and $\eta_c$, respectively. Therefore, we have to consider decays $%
Z_c(3900), \, Z(4430) \to J/\psi \pi; \, \psi^{\prime} \pi$ and $Z_c(3900),
\, Z(4430) \to \eta_c \rho; \, \eta_c^{\prime} \rho $ in the context of QCD
sum rule approach in a correlated form. Reasons behind of such analysis are
clear. In fact, in QCD sum rule method particles are modeled by relevant
interpolating currents which may couple not only to their ground-states, but
also to excited particles.
\begin{table}[tbp]
\begin{tabular}{|c|c|}
\hline\hline
Parameters & Values (in $\mathrm{MeV}$) \\ \hline\hline
$m_{J/\psi}$ & $3096.900 \pm 0.006$ \\
$f_{J/\psi}$ & $411 \pm 7 $ \\
$m_{\psi^{\prime}}$ & $3686.097 \pm 0.025 $ \\
$f_{\psi^{\prime}}$ & $279 \pm 8 $ \\
$m_{\eta_c}$ & $2983.4 \pm 0.5 $ \\
$f_{\eta_c}$ & $404 $ \\
$m_{\eta_{c}^{\prime}}$ & $3686.2 \pm 1.2$ \\
$f_{\eta_{c}^{\prime}}$ & $331$ \\
$m_{\pi}$ & $139.57018 \pm 0.00035$ \\
$f_{\pi}$ & $131.5 $ \\
$m_{\rho}$ & $775.26 \pm 0.25$ \\
$f_{\rho}$ & $216 \pm 3$ \\
$m_{c}$ & $1270 \pm 30$ \\ \hline\hline
\end{tabular}%
\caption{Input parameters.}
\label{tab:Param}
\end{table}

%%%%%%%%%%%%%%%%%%%%%%%%%%%%%%%%%%%%%%%%%%%%%%%%%%%%%%%%%%%%%%%%%%%%%%%%%%%%%%%%%%%

\subsection{$Z_c(3900),\, Z(4430) \to J/\protect\psi \protect\pi; \, \protect%
\psi^{\prime} \protect\pi$ decays}

%%%%%%%%%%%%%%%%%%%%%%%%%%%%%%%%%%%%%%%%%%%%%%%%%%%%%%%%%%%%%%%%%%%%%%%%%%%%%%%%%%%
In order to find width of the decays $Z_{c}(3900)\rightarrow J/\psi \pi
,\,\psi ^{\prime }\pi $ and $Z(4430)\rightarrow J/\psi \pi ;\,\psi ^{\prime
}\pi $ we start from the correlation function
\begin{equation}
\Pi _{\mu \nu }(p,q)=i\int d^{4}xe^{ipx}\langle \pi (q)|\mathcal{T}\{J_{\mu
}^{\psi }(x)J_{\nu }^{Z\dagger }(0)\}|0\rangle ,  \label{eq:CorrF3}
\end{equation}%
where%
\begin{equation}
J_{\mu }^{\psi }(x)=\overline{c}_{i}(x)\gamma _{\mu }c_{i}(x),
\end{equation}%
and $\psi $ denotes one of \ the $J/\psi $ and $\psi ^{\prime }$ mesons. The
interpolating current $J_{\nu }^{Z}(x)$ for the tetraquarks is defined by
Eq.\ (\ref{eq:Curr}).  As we have just emphasized above
the interpolating currents $J_{\nu }^{Z}(x)$ and $J_{\mu }^{\psi }(x)$
couple to $Z_{c}(3900),\,Z(4430)$ and $J/\psi ,\,\psi ^{\prime }$,
respectively. Therefore, the correlation function $\Pi _{\mu \nu }^{\mathrm{%
Phys}}(p,q)$ necessary for our purposes in terms of the mesons' physical
parameters contains four terms
\begin{eqnarray}
&&\Pi _{\mu \nu }^{\mathrm{Phys}}(p,q)=\sum_{\psi =J/\psi ,\psi ^{\prime }}%
\left[ \frac{\langle 0|J_{\mu }^{\psi }|\psi \left( p\right) \rangle }{%
p^{2}-m_{\psi }^{2}}\langle \psi \left( p\right) \pi (q)|Z_{c}(p^{\prime
})\rangle \right.  \notag \\
&&\times \frac{\langle Z_{c}(p^{\prime })|J_{\nu }^{Z\dagger }|0\rangle }{%
p^{\prime 2}-m_{Z_{c}}^{2}}+\frac{\langle 0|J_{\mu }^{\psi }|\psi \left(
p\right) \rangle }{p^{2}-m_{\psi }^{2}}\langle \psi \left( p\right) \pi
(q)|Z(p^{\prime })\rangle  \notag \\
&&\left. \times \frac{\langle Z(p^{\prime })|J_{\nu }^{Z\dagger }|0\rangle }{%
p^{\prime 2}-m_{Z}^{2}}\right] +\dots ,  \label{eq:CorrF4}
\end{eqnarray}%
where the dots stand for contribution of the higher resonances and continuum
states and  $p^{\prime }=p+q$.

To calculate the correlation function we introduce the matrix elements
\begin{eqnarray}
&&\langle 0|J_{\mu }^{\psi }|\psi \left( p\right) \rangle =f_{\psi }m_{\psi
}\varepsilon _{\mu },\,\,\langle Z_{c}(p^{\prime })|J_{\nu }^{Z\dagger
}|0\rangle =f_{Z_{c}}m_{Z_{c}}\varepsilon _{\nu }^{\prime \ast },  \notag \\
&&\langle Z(p^{\prime })|J_{\nu }^{Z\dagger }|0\rangle =f_{Z}m_{Z}\widetilde{%
\varepsilon }_{\nu }^{\prime \ast },  \label{eq:MelA}
\end{eqnarray}%
where $f_{\psi },\,m_{\psi },\,\varepsilon _{\mu }$ are the decay constant,
mass and polarization vector of $J/\psi $ or $\psi ^{\prime }$ meson,
whereas $\widetilde{\varepsilon }_{\nu }^{\prime }$ and $\varepsilon _{\nu
}^{\prime }$ are the polarization vectors of $Z$ and $Z_{c}$ states,
respectively

We need also the vertices
\begin{eqnarray}
\langle \psi \left( p\right) \pi (q)|Z_{c}(p^{\prime })\rangle
&=&g_{Z_{c}\psi \pi }\left[ (p\cdot p^{\prime })(\varepsilon ^{\ast }\cdot
\varepsilon ^{\prime }) \right.  \notag \\
&&\left. -(p\cdot \varepsilon ^{\prime })(p^{\prime }\cdot \varepsilon
^{\ast })\right],  \notag \\
\langle \psi \left( p\right) \pi (q)|Z(p^{\prime })\rangle &=&g_{Z\psi \pi }
\left[ (p\cdot p^{\prime })(\varepsilon ^{\ast }\cdot \widetilde{\varepsilon
}^{\prime })\right.  \notag \\
&&\left. -(p\cdot \widetilde{\varepsilon }^{\prime })(p^{\prime }\cdot
\varepsilon ^{\ast })\right],  \label{eq:MelB}
\end{eqnarray}%
with $g_{Z_{c}\psi \pi }$ and $g_{Z\psi \pi }$ being the strong couplings,
which should be determined from sum rules. By using Eqs.\ (\ref{eq:MelA}) and (%
\ref{eq:MelB}) and after some manipulations we get
\begin{eqnarray}
&&\Pi _{\mu \nu }^{\mathrm{Phys}}(p,q)=\sum_{\psi =J/\psi ,\psi ^{\prime }}%
\left[ \frac{f_{\psi }f_{Z_{c}}m_{Z_{c}}m_{\psi }g_{Z_{c}\psi \pi }}{\left(
p^{\prime 2}-m_{Z_c}^{2}\right) \left( p^{2}-m_{\psi }^{2}\right) }\right.
\notag \\
&&\times \left( \frac{m_{Z_{c}}^{2}+m_{\psi }^{2}}{2}g_{\mu \nu }-p_{\mu
}^{\prime }p_{\nu }\right) +\frac{f_{\psi }f_{Z}m_{Z}m_{\psi }g_{Z\psi \pi }%
}{\left( p^{\prime 2}-m_{Z}^{2}\right) \left( p^{2}-m_{\psi }^{2}\right) }
\notag \\
&&\left. \times \left( \frac{m_{Z}^{2}+m_{\psi }^{2}}{2}g_{\mu \nu }-p_{\mu
}^{\prime }p_{\nu }\right) \right] +\ldots .  \label{eq:CorrF5}
\end{eqnarray}%
For further calculations we choose the structure $\sim g_{\mu \nu }$. The sum of
terms $\sim g_{\mu \nu }$ from Eq.\ (\ref{eq:CorrF5}) constitute the
invariant function which will be used in the following analysis.

The second component of the sum rule is the expression of the same
correlation function given by Eq.\ (\ref{eq:CorrF3}), but computed using
quark propagators. For $\Pi _{\mu \nu }^{\mathrm{QCD}}(p,q)$ we find
\begin{eqnarray}
&&\Pi _{\mu \nu }^{\mathrm{QCD}}(p,q)=\int d^{4}xe^{ipx}\frac{\epsilon
\widetilde{\epsilon }}{\sqrt{2}}\left[ \gamma _{5}\widetilde{S}%
_{c}^{ib}(x){}\gamma _{\mu }\right.  \notag \\
&&\left. \times \widetilde{S}_{c}^{ei}(-x){}\gamma _{\nu }+\gamma _{\nu }%
\widetilde{S}_{c}^{ib}(x){}\gamma _{\mu }\widetilde{S}_{c}^{ei}(-x){}\gamma
_{5}\right] _{\alpha \beta }  \notag \\
&&\times \langle \pi (q)|\overline{u}_{\alpha }^{a}(0)d_{\beta
}^{d}(0)|0\rangle,  \label{eq:CorrF6}
\end{eqnarray}%
where $\alpha $ and $\beta $ are the spinor indices. We continue by
employing the expansion
\begin{equation}
\overline{u}_{\alpha }^{a}d_{\beta }^{d}\rightarrow \frac{1}{4}\Gamma
_{\beta \alpha }^{j}\left( \overline{u}^{a}\Gamma ^{j}d^{d}\right) ,
\label{eq:MatEx}
\end{equation}%
where $\Gamma ^{j}$ is the full set of Dirac matrices
\begin{equation*}
\Gamma ^{j}=\mathbf{1,\ }\gamma _{5},\ \gamma _{\lambda },\ i\gamma
_{5}\gamma _{\lambda },\ \sigma _{\lambda \rho }/\sqrt{2},
\end{equation*}%
As is seen, $\Pi _{\mu \nu }^{\mathrm{QCD}}(p,q)$ instead of the
distribution amplitudes of the pion depends on its local matrix elements.
This is distinctive feature of QCD sum rules on the light-cone when one of
the particles is a tetraquark. As a result, to conserve the four-momentum in
the tetraquark-meson-meson vertex one has to set $q=0$ \cite%
{Agaev:2016dev,Agaev:2017foq}. This restriction should be implemented in the
physical side of the sum rule, as well. In the standard LCSR, $q=0$ is known
as the soft-meson approximation \cite{Belyaev:1994zk}. At vertices composed
of conventional mesons $q\neq 0$, and only in the soft-meson approximation
one equals $q$ to zero, whereas the tetraquark-meson-meson vertex can be
considered in the framework of the LCSR method only if $q=0$. For our
purposes a decisive fact is the observation made in Ref.\ \cite%
{Belyaev:1994zk}: both the soft-meson approximation and full LCSR treatment
of the ordinary mesons' vertices for the strong couplings lead to results
which numerically are very close to each other.

After substituting Eq.\ (\ref{eq:MatEx}) into the expression of the
correlation function and performing the summation over color indices in
accordance with prescriptions presented in a detailed form in Ref.\ \cite%
{Agaev:2016dev}, we determine a local matrix element of the pion that
contribute to $\Pi _{\mu \nu }^{\mathrm{QCD}}(p,q)$, and find the
corresponding spectral density $\rho^{\mathrm{QCD}}(s)$ as its imaginary
part. It turns out that only the local matrix element of the pion
\begin{equation}
\langle 0|\overline{d}(0)i\gamma _{5}u(0)|\pi (q)\rangle = f_{\pi }\mu_{\pi }
\label{eq:MatE2}
\end{equation}%
contribute to $\rho^{\mathrm{QCD}}(s)$, where $\mu_{\pi}=m_{\pi
}^{2}/(m_{u}+m_{d})$.

To calculate $\rho ^{\mathrm{QCD}}(s)$ we choose in $\Pi _{\mu \nu }^{%
\mathrm{QCD}}(p,q)$ again the structure $\sim g_{\mu \nu }$, and get
\begin{equation}
\rho ^{\mathrm{QCD}}(s)=\frac{f_{\pi }\mu _{\pi }}{12\sqrt{2}}\left[ \rho ^{%
\mathrm{pert.}}(s)+\rho ^{\mathrm{n.-pert.}}(s)\right],  \label{eq:SD1}
\end{equation}%
where $\rho ^{\mathrm{pert.}}(s)$ and $\rho ^{\mathrm{n.-pert.}}(s)$ are the
perturbative and nonperturbative components of the spectral density. The
perturbative part of $\rho ^{\mathrm{QCD}}(s)$ has a rather simple form and
was calculated in Ref.\ \cite{Agaev:2016dev}:
\begin{equation}
\rho ^{\mathrm{pert.}}(s)=\frac{(s+2m_{c}^{2})\sqrt{s(s-4m_{c}^{2})}}{\pi ^{2}s%
}.
\end{equation}%
The $\rho ^{\mathrm{n.-pert.}}(s)$ contains nonperturbative contributions $%
\sim \langle \alpha _{s}G^{2}/\pi \rangle $, $\sim \langle
g_{s}^{3}G^{3}\rangle $ and $\sim \langle \alpha _{s}G^{2}/\pi \rangle ^{2}$
which are terms of four, six and eight dimensions, respectively. Stated
differently, $\rho ^{\mathrm{n.-pert.}}(s)$ encompasses nonperturbative
contributions up to eight dimensions: its explicit expression is moved to
Appendix.

Having found $\rho ^{\mathrm{QCD}}(s)$ which constitutes the QCD side of the
desiring sum rule, and clarified the necessity of the limit $q\rightarrow 0$%
, we turn back to finish calculation of its phenomenological side. In the
limit $q\rightarrow 0$ we get $p^{\prime }=p$ and invariant function
corresponding to the structure $g_{\mu \nu }$ in Eq.\ (\ref{eq:CorrF4})
depends only on the variable $p^{2}$ and has the form
\begin{eqnarray}
&&\Pi ^{\mathrm{Phys}}(p^{2})=\frac{f_{J/\psi }f_{Z_{c}}m_{Z_{c}}m_{J/\psi
}m_{1}^{2}}{\left( p^{2}-m_{1}^{2}\right) ^{2}}g_{Z_{c}J/\psi \pi }  \notag
\\
&&+\frac{f_{\psi ^{\prime }}f_{Z_{c}}m_{Z_{c}}m_{\psi ^{\prime }}m_{2}^{2}}{%
\left( p^{2}-m_{2}^{2}\right) ^{2}}g_{Z_{c}\psi ^{\prime }\pi }+\frac{%
f_{J/\psi }f_{Z}m_{Z}m_{J/\psi }m_{3}^{2}}{\left( p^{2}-m_{3}^{2}\right) ^{2}%
}g_{ZJ/\psi \pi }  \notag \\
&&+\frac{f_{\psi ^{\prime }}f_{Z}m_{Z}m_{\psi ^{\prime }}m_{4}^{2}}{\left(
p^{2}-m_{4}^{2}\right) ^{2}}g_{Z\psi ^{\prime }\pi }+\ldots ,
\label{eq:PhysSide1}
\end{eqnarray}%
where $m_{1}^{2}=(m_{Z_{c}}^{2}+m_{J/\psi }^{2})/2$, $%
m_{2}^{2}=(m_{Z_{c}}^{2}+m_{\psi ^{\prime }}^{2})/2$,$\
m_{3}^{2}=(m_{Z}^{2}+m_{J/\psi }^{2})/2$, and $m_{4}^{2}=(m_{Z}^{2}+m_{\psi
^{\prime }}^{2})/2$.

In the limit $q\rightarrow 0$ the phenomenological side of the sum rules
apart from the strong coupling of the ground-state particles $g_{Z_{c}J/\psi
\pi }$ contains also other terms which are not suppressed relative to the
main term even after the Borel transformation \cite{Belyaev:1994zk}. In
order to eliminate their effects the operator
\begin{equation}
\mathcal{P}(M^{2},m^{2})=\left( 1-M^{2}\frac{d}{dM^{2}}\right)
M^{2}e^{m^{2}/M^{2}},  \label{eq:softop}
\end{equation}%
should be applied to both sides of sum rules \cite{Ioffe:1983ju}. In our
previous works devoted to investigation of tetraquarks for calculation of
their strong couplings and decay widths we applied namely this technique
(see, for example Refs.\ \cite{Agaev:2016dev,Agaev:2017foq}). But these
terms arise from vertices composed of excited states of the initial (final)
particles, i.e. in our case from $ZJ/\psi \pi $, and $Z_{c}\psi ^{\prime
}\pi $ vertices. Stated differently, unsuppressed terms treated as
contamination when studying vertices of ground-state particles now are
subject of investigation. Because $\Pi ^{\mathrm{Phys}}(p^{2})$ is a sum of
four terms, and even at the first phase of calculation, which will be
explained below, contains at least two of them, we are not going to apply
the operator $\mathcal{P}$ to these expressions.

To proceed we follow the recipe used in the previous section, i.e. we choose
the parameter $s_{0}$ below threshold of the $Z\rightarrow J/\psi \pi $ and $%
Z\rightarrow \psi ^{\prime }\pi $ decays. Then in the explored range of $%
s\in (0,s_{0})$ in Eq.\ (\ref{eq:PhysSide1}) only first two terms have to be
taken into account explicitly: last two terms are, naturally, included into
a "higher resonances and continuum". Applying the one-variable Borel
transformation to the survived terms, equating the physical and QCD sides of
the sum rule, and performing the continuum subtraction in accordance with
hadron-quark duality we derive the following expression
\begin{eqnarray}
&&f_{Z_{c}}m_{Z_{c}} \left[f_{J/\psi }m_{J/\psi }m_{1}^{2}g_{Z_{c}J/\psi \pi
}e^{-m_{1}^{2}/M^{2}} +f_{\psi^{\prime} }m_{\psi^{\prime} }m_{2}^{2} \right.
\notag \\
&&\left. \times g_{Z_{c}\psi^{\prime} \pi }e^{-m_{2}^{2}/M^{2}}\right]
=M^2\int_{4m_{c}^{2}}^{s_{0}}dse^{-s/M^{2}}\rho ^{\mathrm{QCD}}(s).
\label{eq:SR1}
\end{eqnarray}%
But only this equality is not enough to extract two couplings $g_{Z_{c}\psi
^{\prime }\pi }$ and $g_{Z_{c}J/\psi \pi }$. The second expression is
derived by applying the operator $d/d(-1/M^{2})$ to both sides of Eq.\ (\ref%
{eq:SR1}). The obtained expression in conjunction with Eq.\ (\ref{eq:SR1})
allows us to derive sum rules for these two couplings. They will be used to
calculate width of the decays $Z_{c}\rightarrow \psi ^{\prime }\pi $ and $%
Z_{c}\rightarrow J/\psi \pi $, and will enter as input parameters to the
next sum rules.

The next two sum rules are obtained by fixing $\sqrt {s_0^{\star}}%
=m_Z+(0.3-0.7)~\mathrm{GeV}$. The choice for $s_0^{\star}$ is motivated by
observation that a mass splitting in the tetraquark multiplet may be $\sim
(0.3-0.7)~\mathrm{GeV}$. For $s\in (0,s_{0}^{\star})$ the processes $Z\to
J/\psi \pi $ and $Z \to \psi^{\prime }\pi $ have to be taken into account.
In other words, in this phase of analysis all of four terms in Eq.\ (\ref%
{eq:PhysSide1}) should be considered explicitly: while two of them are
known, we have to extract remaining couplings $g_{Z\psi^{\prime }\pi }$ and $%
g_{ZJ/\psi \pi }$. To this end, we repeat operations described above and
derive last two sum rules for the required couplings.

The width of the decay $Z \to \psi \pi$, where $Z=Z(4430)$ or $Z_c(3900)$,
and $\psi = J/\psi;\, \psi^{\prime}$ can be evaluated by means of the
formula
\begin{eqnarray}
&&\Gamma \left( Z\to \psi \pi\right) =\frac{g_{Z\psi \pi }^{2}m_{\psi }^{2}}{%
24\pi }\lambda \left( m_{Z},\ m_{\psi },m_{\pi }\right)  \notag \\
&&\times \left[ 3+\frac{2\lambda ^{2}\left( m_{Z},\ m_{\psi },m_{\pi
}\right) }{m_{\psi }^{2}}\right] ,  \label{eq:DW}
\end{eqnarray}%
where
\begin{equation*}
\lambda (a,\ b,\ c)=\frac{\sqrt{a^{4}+b^{4}+c^{4}-2\left(
a^{2}b^{2}+a^{2}c^{2}+b^{2}c^{2}\right) }}{2a}.
\end{equation*}

As is seen, besides the strong coupling $g_{Z\psi \pi}$ the decay width
depends also on the parameters of the tetraquark and final mesons. The mass
and current coupling of $Z(4430)$ and $Z_c(3900)$ resonances are calculated
in the present work. The mass of the $J/\psi$, $\psi^{\prime}$, $\pi$, as
well as $\eta_c$, $\eta_c^{\prime}$ and $\rho$ mesons which will be used in
the next subsection can be found in Ref.\ \cite{Olive:2016xmw}. For the
decay constants $f_{J/\psi} $ and $f_{\psi^{\prime}}$ we use the same
source, whereas $f_{\eta_c} $ and $f_{\eta_c^{\prime}}$ are borrowed from
Ref.\ \cite{Negash:2015rua}: All these information are collected in Table\ %
\ref{tab:Param}.

In numerical computations we have used the same ranges for the Borel
parameter and $s_{0}$ as in mass and current coupling analysis. Another
question to be addressed here is contribution of the "excited" terms to the
sum rules. It is known, that the ground-state contributes dominantly to the
spectral density. But besides the strong coupling of the ground-state
particles, we extract also couplings of the vertices, where one or two of
particles are radially excited states. Their contributions to the sum rules
should be sizeable to lead reliable predictions for evaluating quantities.
To check this point we calculate the pole contribution to the sum rules
defined as
\begin{equation}
\mathrm{PC}=\frac{\int_{0}^{s_{0}}ds\rho ^{\mathrm{QCD}}(s)e^{-s/M^{2}}}{%
\int_{0}^{\infty }ds\rho ^{\mathrm{QCD}}(s)e^{-s/M^{2}}}.  \label{eq:PC}
\end{equation}%
Choosing $s_{0}=4.2^{2}\ \mathrm{GeV}^{2}$ and fixing $M^{2}=4.5\ \mathrm{GeV%
}^{2}$ we get $\mathrm{PC}=0.81$, which is formed due to terms $\sim
g_{Z_{c}J/\psi \pi }$ and $\sim g_{Z_{c}\psi ^{\prime }\pi }$. At the next
stage, we fix $s_{0}\equiv s_{0}^{\star }$ and find $\mathrm{PC}=0.95$,
which now contain contribution coming from four terms. This means that the
excited terms $\sim g_{ZJ/\psi \pi }$ and $\sim g_{Z\psi ^{\prime }\pi }$
constitute approximately $14\%$ part of the sum rules. From this analysis,
one can see, first of all, that the working window for the Borel parameter
is found correctly, because $\mathrm{PC}>1/2$ is one of the constraints in
fixing of $M^{2}$. Secondly, effects of the terms related directly to $Z$
decays are numerically small, nevertheless couplings $g_{ZJ/\psi \pi }$ and $%
g_{Z\psi ^{\prime }\pi }$ are computed using expressions, which contain
contributions all of terms, and hence evaluating of the strong couplings are
based on reliable sum rules. Finally, an effect of the "higher excited
states and continuum" does not exceed $5\%$ of $\mathrm{PC}$, which a
posteriori confirms our suggestion tacitly made in writing the sum rule (\ref%
{eq:SR1}), which implies that a contamination of the physical side by
excited states higher than $Z(4430)$ resonance is negligible.

\begin{widetext}

\begin{figure}[h!]
\begin{center}
\includegraphics[totalheight=6cm,width=8cm]{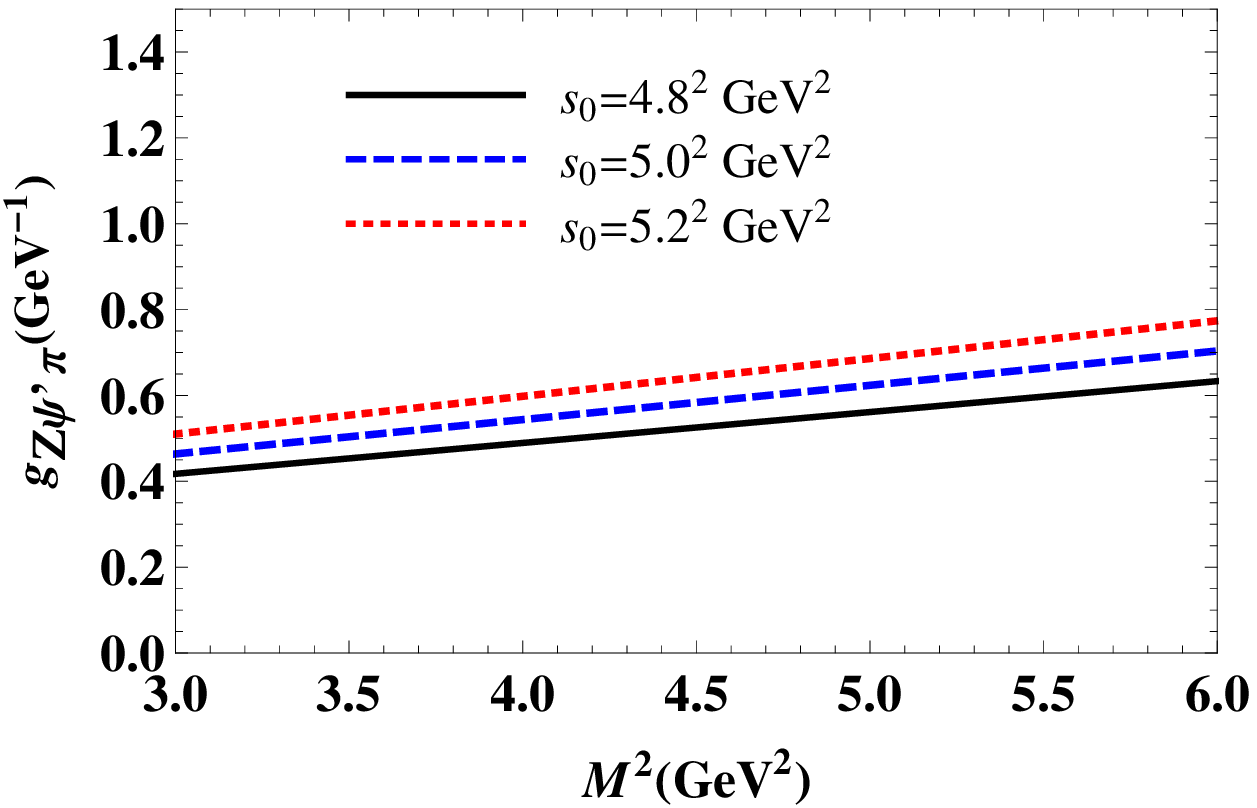}\,\, %
\includegraphics[totalheight=6cm,width=8cm]{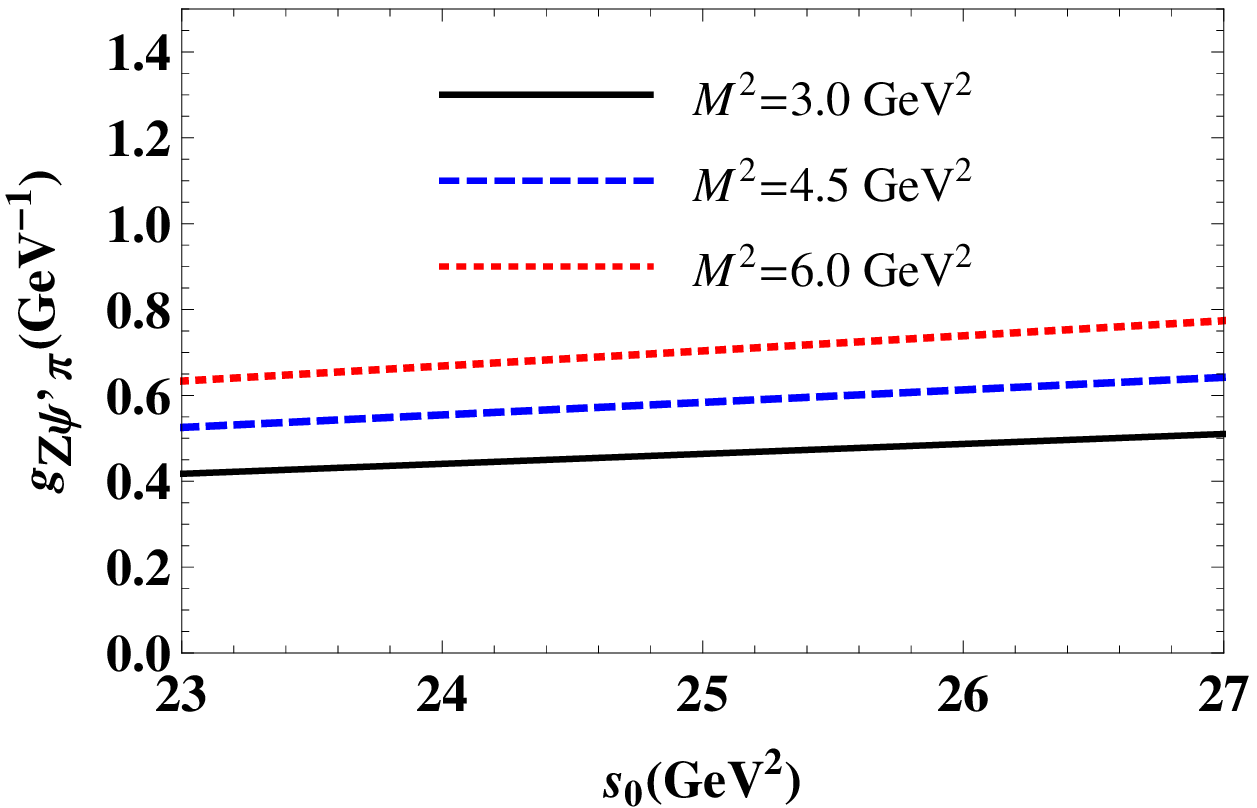}
\end{center}
\caption{ The coupling $g_{Z\psi^{\prime}\pi}$  as a function of the Borel parameter
$M^2$ at fixed $s_0$ (left panel), and as a function of the continuum threshold
$s_0$ at fixed $M^2$ (right panel).}
\label{fig:GZPsiPi}
\end{figure}

\end{widetext}

The couplings $g$ depend on $M^2$ and $s_0$ remaining nevertheless within
limits which are typical for such kind of calculations. This variation of
the couplings together with uncertainties coming from other parameters
generates final theoretical errors of numerical analysis. To visualize these effects
we plot in Fig.\ \ref{fig:GZPsiPi} , as an example, the dependence of the coupling
$g_{Z\psi^{\prime}\pi}$ on the parameters $M^2$ and $s_0$.
For the strong coupling $g_{Z\psi^{\prime}\pi }$ and width of the
corresponding decay $Z\rightarrow \psi^{\prime}\pi $ process numerical
computations predict:
\begin{eqnarray}
&&g_{Z\psi^{\prime}\pi } =(0.58 \pm 0.17)\ \mathrm{GeV}^{-1},  \notag \\
&&\Gamma(Z\rightarrow \psi^{\prime}\pi )=(129.7 \pm 38.4)\ \mathrm{MeV}.
\label{eq:PiDW1}
\end{eqnarray}
The coupling $g_{Z J/\psi \pi}$ and width of the decay $Z \to J/\psi \pi$
are found as
\begin{eqnarray}
&&g_{Z J/\psi \pi} =(0.24 \pm 0.07) \ \mathrm{GeV}^{-1},  \notag \\
&&\Gamma (Z \to J/\psi \pi )=(27.4 \pm 7.3)\ \mathrm{MeV}.  \label{eq:PiDW2}
\end{eqnarray}
Our results for all of four strong couplings, as well as the decay width of
corresponding processes are collected in Table\ \ref{tab:Results1A}.
\begin{table}[tbp]
\begin{tabular}{|c|c|c|c|c|}
\hline
Channels & $Z \to \psi^{\prime} \pi$ & $Z \to J/\psi \pi$ & $Z_c \to
\psi^{\prime} \pi$ & $Z_c \to J/\psi \pi$ \\ \hline\hline
$g ~(\mathrm{GeV}^{-1})$ & $0.58 \pm 0.17$ & $0.24 \pm 0.07 $ & $0.29 \pm
0.08$ & $0.38 \pm 0.11$ \\ \hline
$\Gamma ~(\mathrm{MeV})$ & $129.7 \pm 38.4$ & $27.4 \pm 7.3$ & $7.1 \pm 1.9$
& $39.9 \pm 9.3$ \\ \hline\hline
\end{tabular}%
\caption{The strong coupling $g$ and width $\Gamma$ of the $Z(Z_c) \to
\protect\psi^{\prime}(J/\protect\psi)\protect\pi$ decay channels.}
\label{tab:Results1A}
\end{table}

%%%%%%%%%%%%%%%%%%%%%%%%%%%%%%%%%%%%%%%%%%%%%%%%%%%%%%%%%%%%%%%%%%%%%%%%%%%%%%%%%%%

\subsection{$Z_c(3900),\, Z(4430) \to \protect\eta_c^{\prime}\protect\rho; \,%
\protect\eta_c \protect\rho$ decays}

%%%%%%%%%%%%%%%%%%%%%%%%%%%%%%%%%%%%%%%%%%%%%%%%%%%%%%%%%%%%%%%%%%%%%%%%%%%%%%%%%%%
The $Z_c(3900)$ and $Z(4430)$ may decay also to final states $\eta_c \rho$
and $\eta_c^{\prime} \rho$. Here we have three decay modes $Z \to
\eta_c^{\prime} \rho$, $Z \to \eta_c \rho$ and $Z_c \to \eta_c \rho$: the
process $Z_c \to \eta_c^{\prime} \rho$ is forbidden kinematically. An
analysis of three decay channels in some points differs from similar studies
fulfilled in the previous subsection.

As usual, we consider the correlation function
\begin{equation}
\Pi _{\nu }(p,q)=i\int d^{4}xe^{ipx}\langle \rho (q)|\mathcal{T}\{J^{\eta
_{c}}(x)J_{\nu }^{Z\dagger }(0)\}|0\rangle ,
\end{equation}%
where $\eta _{c}\equiv \eta _{c},\ \eta _{c}^{\prime }$, and the current $%
J^{\eta _{c}}(x)$ is defined as%
\begin{equation*}
J^{\eta _{c}}(x)=\overline{c}_{i}(x)i\gamma _{5}c_{i}(x).
\end{equation*}%
In order to find the hadronic representation of the correlation function we
define the matrix elements
\begin{equation}
\langle 0|J^{\eta _{c}}|\eta _{c}(p)\rangle =\frac{f_{\eta _{c}}m_{\eta
_{c}}^{2}}{2m_{c}},
\end{equation}%
with $m_{\eta _{c}}$ and $f_{\eta _{c}}$ being the $\eta _{c}$ ($\eta
_{c}^{\prime }$) mesons' mass and decay constant. The relevant vertices are
introduced in the following form
\begin{eqnarray}
\langle \eta _{c}\left( p\right) \rho (q)|Z(p^{\prime })\rangle &=&g_{Z\eta
_{c}\rho }\left[ (q\cdot \widetilde{\varepsilon }^{\prime })(p^{\prime
}\cdot \varepsilon ^{\ast })\right.  \notag \\
&&-\left. (q\cdot p^{\prime })(\varepsilon ^{\ast }\cdot \widetilde{%
\varepsilon }^{\prime })\right] ,
\end{eqnarray}
and
\begin{eqnarray}
\langle \eta _{c}\left( p\right) \rho (q)|Z_{c}(p^{\prime })\rangle &=&
g_{Z_{c}\eta _{c}\rho } \left[ (q\cdot \varepsilon ^{\prime })(p^{\prime
}\cdot \varepsilon ^{\ast })\right.  \notag \\
&&-\left. (q\cdot p^{\prime })(\varepsilon ^{\ast }\cdot \varepsilon
^{\prime })\right] ,
\end{eqnarray}%
with $q$ and $\varepsilon $ being the momentum and polarization vector of
the $\rho $-meson, respectively.

Then the calculation of the hadronic representation $\Pi _{\nu }^{\mathrm{%
Phys}}(p,q)$ is straightforward and yields
\begin{eqnarray}
&&\Pi _{\nu }^{\mathrm{Phys}}(p,q)=\frac{\langle 0|J^{\eta _{c}}|\eta
_{c}\left( p\right) \rangle }{p^{2}-m_{\eta _{c}}^{2}}\langle \eta
_{c}\left( p\right) \rho (q)|Z_{c}(p^{\prime })\rangle  \notag \\
&&\times \frac{\langle Z_{c}(p^{\prime })|J_{\nu }^{Z}|0\rangle }{p^{\prime
2}-m_{Z_{c}}^{2}}+\sum_{\eta _{c}=\eta_{c}, \eta _{c}^{\prime}}\frac{\langle
0|J^{\eta _{c}}|\eta _{c}\left( p\right) \rangle }{p^{2}-m_{\eta _{c}}^{2}}%
\langle \eta _{c}\left( p\right) \rho (q)|Z(p^{\prime })\rangle  \notag \\
&&\times \frac{\langle Z(p^{\prime })|J_{\nu }^{Z}|0\rangle }{p^{\prime
2}-m_{Z}^{2}}+\dots .
\end{eqnarray}%
Employing the required matrix elements, for the invariant amplitude
corresponding to the structure $\sim \epsilon _{\nu }^{\ast }$ in the limit $%
q\rightarrow 0$ we find
\begin{eqnarray}
&&\Pi ^{\mathrm{Phys}}(p^{2})=\frac{f_{\eta _{c}}f_{Z_{c}}m_{Z_{c}}m_{\eta
_{c}}^{2}g_{Z_{c}\eta _{c}\rho }}{4m_{c}\left( p^{2}-\widetilde{m}%
_{1}^{2}\right) ^{2}}(m_{Z_{c}}^{2}-m_{\eta _{c}}^{2})  \notag \\
&&+\frac{f_{\eta _{c}}f_{Z}m_{Z}m_{\eta _{c}}^{2}g_{Z\eta _{c}\rho }}{%
4m_{c}\left( p^{2}-\widetilde{m}_{2}^{2}\right) ^{2}}(m_{Z}^{2}-m_{\eta
_{c}}^{2})  \notag \\
&&+\frac{f_{\eta _{c}^{\prime }}f_{Z}m_{Z}m_{\eta _{c}^{\prime
}}^{2}g_{Z\eta _{c}^{\prime }\rho }}{4m_{c}\left( p^{2}-\widetilde{m}%
_{3}^{2}\right) ^{2}}(m_{Z}^{2}-m_{\eta _{c}^{\prime }}^{2})+\ldots ,
\label{eq:PhysSide}
\end{eqnarray}%
where the notations $\widetilde{m}_{1}^{2}=(m_{Z_{c}}^{2}+m_{\eta
_{c}}^{2})/2$, $\widetilde{m}_{2}^{2}=(m_{Z}^{2}+m_{\eta _{c}}^{2})/2$ and $%
\widetilde{m}_{3}^{2}=(m_{Z}^{2}+m_{\eta _{c}^{\prime }}^{2})/2$ are
introduced.

Computation of the correlation function $\Pi _{\nu }^{\mathrm{QCD}}(p,q)$
using quark propagators yields
\begin{eqnarray}
&&\Pi _{\nu }^{\mathrm{QCD}}(p,q)=-i\int d^{4}xe^{ipx}\frac{\epsilon
\widetilde{\epsilon }}{\sqrt{2}}\left[ \gamma _{5}\widetilde{S}%
_{c}^{ib}(x){}\gamma _{5}\right.  \notag \\
&&\left. \times \widetilde{S}_{c}^{ei}(-x){}\gamma _{\nu }+\gamma _{\nu }%
\widetilde{S}_{c}^{ib}(x){}\gamma _{5}\widetilde{S}_{c}^{ei}(-x){}\gamma _{5}%
\right] _{\alpha \beta }  \notag \\
&&\times \langle \rho (q)|\overline{u}_{\alpha }^{d}(0)d_{\beta
}^{a}(0)|0\rangle .  \label{eq:CorrF7}
\end{eqnarray}%
In the $q \to 0$ limit only the matrix elements
\begin{equation}
\langle 0|\overline{u}(0)\gamma _{\mu }d(0)|\rho (p,\lambda )\rangle
=\epsilon _{\mu }^{(\lambda )}f_{\rho }m_{\rho },  \label{eq:Melem1}
\end{equation}
and
\begin{equation}
\langle 0|\overline{u}(0)g\widetilde{G}_{\mu \nu }\gamma _{\nu }\gamma
_{5}d(0)|\rho (p,\lambda )\rangle =f_{\rho }m_{\rho }^{3}\epsilon _{\mu
}^{(\lambda )}\zeta _{4\rho} ,  \label{eq:Melem2}
\end{equation}
contribute to the spectral density $\rho^{\mathrm{QCD}}(s)$ \cite%
{Agaev:2016dev}. They depend on the mass and decay constant of the $\rho$%
-meson $m_{\rho}$, $f_{\rho}$, and on $\zeta _{4\rho}$ which normalizes the
twist-4 matrix element of the $\rho$-meson \cite{Ball:1998ff}. The parameter
$\zeta _{4\rho}$ was evaluated in the context of QCD sum rule approach at
the renormalization scale $\mu =1\,\,{\mathrm{GeV}}$ in Ref.\ \cite%
{Ball:2007zt} and is equal to $\zeta _{4\rho }=0.07\pm 0.03$.

The spectral density $\rho^{\mathrm{QCD}}(s)$ is derived in accordance with
known recipes and is given as
\begin{equation}
\rho ^{\mathrm{QCD}}(s)=\frac{f_{\rho } m_{\rho }}{8\sqrt{2}}\left[ \rho ^{%
\mathrm{pert.}}(s)+\rho ^{\mathrm{n.-pert.}}(s)\right],  \label{eq:SD2}
\end{equation}
where
\begin{equation}
\rho ^{\mathrm{pert.}}(s)=\frac{\sqrt{s(s-4m_{c}^{2})}}{\pi ^{2}}.
\end{equation}
The nonperturbative component of $\rho ^{\mathrm{QCD}}(s)$ is calculated
with dimension-8 accuracy: its explicit form can be found in Appendix.

On order to derive sum rules we use an iterative approach explained in the
previous subsection. At the first stage our task is simple. Really, for $%
s\in (0,\ s_{0})$ there is only one term in the physical side of the sum
rule. At this step we evaluate only the ground-state coupling $g_{Z_{c}\eta
_{c}\rho }$, therefore can apply the operator $\mathcal{P}$ to clean the
physical side of the sum rule form effects of excited states. As a result,
we get%
\begin{eqnarray*}
g_{Z_{c}\eta _{c}\rho } &=&\frac{4m_{c}}{f_{\eta
_{c}}f_{Z_{c}}m_{Z_{c}}m_{\eta _{c}}^{2}(m_{Z_{c}}^{2}-m_{\eta _{c}}^{2})} \\
&&\times \mathcal{P}(M^{2},\widetilde{m}_{1}^{2})%
\int_{4m_{c}^{2}}^{s_{0}}dse^{-s/M^{2}}\rho ^{\mathrm{QCD}}(s).
\end{eqnarray*}%
In the domain $s\in (0,\ s_{0}^{\ast })$ all terms in Eq.\ (\ref{eq:PhysSide}%
) become active, and we obtain the expression containing two remaining
unknown couplings. Because excited terms enter to this expression explicitly
and our aim is to find corresponding couplings, we do not apply the operator
$\mathcal{P}$. The system of equations can be completed by using the
operator $d/d(-1/M^{2})$ to both sides of this expression which leads the
second equality. Solutions of the obtained equations are sum rules for the
couplings $g_{Z\eta _{c}}\rho $ and $g_{Z\eta _{c}}^{\prime }\rho $. The
width of the decays $Z\rightarrow \eta _{c}\rho $, $Z\rightarrow \eta
_{c}^{\prime }\rho $ and $Z_{c}\rightarrow \eta _{c}\rho $ can be calculated
by means of Eq.\ (\ref{eq:DW}) with replacements $m_{\pi }\rightarrow
m_{\eta _{c}}(m_{\eta _{c}^{\prime }})$ and $m_{\psi }\rightarrow m_{\rho }$.

For the coupling $g_{Z_{c}\eta _{c}\rho }$ and width of the decay $%
Z_{c}\rightarrow \eta _{c}\rho $ we get
\begin{eqnarray}
&&g_{Z_{c}\eta _{c}\rho } =(1.28 \pm 0.32)\ \mathrm{GeV}^{-1},  \notag
\\
&&\Gamma (Z_{c} \rightarrow \eta _{c}\rho )=(20.28 \pm 5.17)\ \mathrm{%
MeV}.  \label{eq:RhoDW1}
\end{eqnarray}
For the strong couplings $g_{Z\eta _{c}^{\prime}\rho }$ and $g_{Z\eta
_{c}\rho }$, and width of the processes $Z \to \eta _{c}^{\prime}\rho$ and $%
Z \to \eta _{c}\rho$ we find
\begin{eqnarray}
&&g_{Z\eta _{c}^{\prime}\rho } =(0.81 \pm 0.21) \ \mathrm{GeV}^{-1},
\notag \\
&&\Gamma (Z \to \eta _{c}^{\prime}\rho )=(1.01 \pm 0.28)\ \mathrm{MeV},
\label{eq:RhoDW2}
\end{eqnarray}
and
\begin{eqnarray}
&&g_{Z\eta _{c}\rho } =(0.48 \pm 0.13)\ \mathrm{GeV}^{-1},  \notag \\
&&\Gamma (Z \to \eta _{c}\rho )=(11.57 \pm 3.04)\ \mathrm{MeV}.
\label{eq:RhoDW3}
\end{eqnarray}

The decays $Z_c \to J/\psi \pi$ and $Z_c \to \eta_c \rho$ were considered in
our previous work \cite{Agaev:2016dev} using QCD sum rules on light-cone and
diquark-antidiquark interpolating current. In Table\ \ref{tab:ZcDec} the
partial decay width of these channels obtained in Ref.\ \cite{Agaev:2016dev}
are compared with predictions of the present investigation. As is seen, they
do not differ considerably from each other: it is remarkable, that iterative
scheme adopted in this work lead to almost identical to Ref.\ \cite%
{Agaev:2016dev} predictions, which may be considered as serious
consistency-check of the employed approach. The small discrepancies between
two sets of predictions can be attributed to accuracy of the spectral densities,
which in the present work have been found by taking into account condensates up to eight
dimensions, whereas in Ref.\ \cite{Agaev:2016dev} $\rho_{\pi}^{\mathrm{QCD}%
}(s)$ and $\rho_{\rho}^{\mathrm{QCD}}(s)$ contained only perturbative terms.
It is worth noting that, in the present work we have evaluated the partial
width of the $Z_c \to \psi^{\prime} \pi$ mode, which was not considered in
Ref.\ \cite{Agaev:2016dev}.

Our results for the decays of the $Z(4430)$ resonance are collected in
Table\ \ref{tab:ZDec}. As is seen, $Z$ dominantly decays through $Z \to
\psi^{\prime} \pi$ channel. The sum of predictions for the channels $Z \to
\psi^{\prime} \pi$ and $Z \to J/\psi \pi$ $(157.1\pm 39.1)\ {\mathrm{MeV}}$
agrees with LHCb measurements (see, Eq.\ (\ref{eq:LHCdata})) staying below
the upper limit of experimental data $\sim 212\ {\mathrm{MeV}}$.
Unfortunately, an experimental information on the decay width $\Gamma(Z \to
J/\psi \pi)$ is restricted by Belle report on product of branching fractions
$\mathcal{B}(\bar{B}^{0} \to K^{-}Z(4430)^{+})\mathcal{B}(Z(4430)^{+} \to
J/\psi \pi)=(5.4^{+4.0}_{-1.0}{}^{+1.1}_{-0.9})\cdot 10^{-6} $. It is
possible, by invoking a similar experimental measurements for $\psi^{\prime}$
to estimate a ratio
\begin{equation}
R_Z=\Gamma(Z \to \psi^{\prime} \pi)/\Gamma(Z \to J/\psi \pi),
\end{equation}
which actually was done in Ref.\ \cite{Goerke:2016hxf}. But we are not going
to make far-reaching conclusions from similar estimates: From our point of
view, in the lack of direct measurements of $\Gamma(Z \to J/\psi \pi)$, the
best what can be done is calculation of theoretical prediction for $R_Z$,
which equals in our case to $R_Z=4.73 \pm 1.88$.
\begin{table}[t]
\begin{tabular}{|c|c|c|c|}
\hline\hline
& $\Gamma(Z_c \to J/ \psi \pi)$ & $\Gamma(Z_c \to \psi^{\prime} \pi)$ & $%
\Gamma(Z_c \to \eta_c \rho)$  \\
& $(\mathrm{MeV})$ & $(\mathrm{MeV})$ & $(\mathrm{MeV})$   \\ \hline
This work & $39.9 \pm 9.3$ & $7.1 \pm 1.9$ & $20.3 \pm 5.2$   \\
\hline
\cite{Agaev:2016dev} & $41.9\pm 9.4$ & $-$ & $23.8\pm 4.9$   \\ \hline
\cite{Dias:2013xfa} & $29.1 \pm 8.2$ & $-$ & $27.5\pm 8.5$  \\ \hline
\cite{Goerke:2016hxf}A & $27.9^{+6.3}_{-5.0}$ & $-$ & $35.7^{+6.3}_{-5.2}$
\\ \hline
\cite{Goerke:2016hxf}B & $1.8 \pm 0.3$ & $-$ & $3.2^{+0.5}_{-0.4}$   \\
\hline
\cite{Dong:2013iqa} & $10.43 - 23.89$ & $1.28-2.94$ & $-$   \\ \hline\hline
\end{tabular}%
\caption{Theoretical results for some of the $Z_c(3900)$ resonance's decay
modes.}
\label{tab:ZcDec}
\end{table}
\begin{table}[t]
\begin{tabular}{|c|c|c|c|c|}
\hline\hline
& $\Gamma(Z\to J/ \psi \pi)$ & $\Gamma(Z \to \psi^{\prime} \pi)$ & $\Gamma(Z
\to \eta_c \rho)$ & $\Gamma(Z \to \eta_c^{\prime} \rho)$   \\
& $(\mathrm{MeV})$ & $(\mathrm{MeV})$ & $(\mathrm{MeV})$ & $(\mathrm{MeV})$ \\ \hline
Th. w. & $27.4 \pm 7.3$ & $129.7 \pm 38.4$ & $11.6 \pm 3.0$ & $%
1.0 \pm 0.3$   \\ \hline
\cite{Goerke:2016hxf} & $26.9$ & $120.6$ & $-$ & $-$   \\ \hline\hline
\end{tabular}%
\caption{The same as in Table\ \protect\ref{tab:ZcDec}, but for the $Z(4430)$
state.}
\label{tab:ZDec}
\end{table}

\section{Summary and Concluding notes}

The decays of $Z$ and $Z_c$ resonances were previously investigated in Ref.\
\cite{Goerke:2016hxf,Dias:2013xfa,Dong:2013iqa}: some of their results are
shown in Tables\ \ref{tab:ZcDec} and \ref{tab:ZDec}. In Ref.\ \cite%
{Dias:2013xfa} using the three-point sum rule method and diquark-antidiquark
interpolating current authors calculated partial decay width of the channels
$Z_c \to J/\psi \pi$, $Z_c \to \eta_c \rho$, $Z_c \to \bar{D}^{0}D^{\star}$
and $Z_c \to \bar{D}^{\star 0}D$ . Results for two first modes can be found
in Table\ \ref{tab:ZcDec}.

Within the covariant quark model decays of the $Z_c(3900)$ state were
analyzed in Ref.\ \cite{Goerke:2016hxf}, where it was considered both as a
diqaurk-antidiquak and molecule-type tetraquarks. Thus, by treating $Z_c$ as
a four-quark system with diquark-antidiquark composition and using a size
parameter $\Lambda_{Z_c}=2.25 \pm 0.10\ \mathrm{GeV}$ in their model (model
A) authors evaluated width of the decays $Z_c \to J/\psi \pi$, $Z_c \to
\eta_c \rho$ (see, Table\ \ref{tab:ZcDec}). Assuming a molecular-type
structure for $Z_c(3900)$ and choosing $\Lambda_{Z_c}=3.3 \pm 0.1\ \mathrm{%
GeV}$ (model B) the same decay widths were calculated in Ref.\ \cite%
{Goerke:2016hxf}: obtained predictions are shown in Table\ \ref{tab:ZcDec},
too.

The decays of $Z_c(3900)$ state in the framework of a phenomenological
Lagrangian approach were considered in Ref.\ \cite{Dong:2013iqa}. The $Z_c$
state there was treated as a hadronic molecule composed of $\bar{D}D^{\star}$
and $\bar{D}^{\star}D$. For a binding energy of the hadronic molecule $%
\epsilon=20\ \mathrm{MeV}$ authors found the width of different decay
channels, some of which are demonstrated in Table\ \ref{tab:ZcDec}.

Decays of the $Z(4430)$ resonance to $J/\psi \pi$ and $\psi^{\prime}\pi$ in
the context of the diquark-antidiquark model were studied in Ref.\ \cite%
{Goerke:2016hxf}. Predictions for the partial width of these decays (see,
Table\ \ref{tab:ZDec}) obtained at $\Lambda_{Z(4430)}=2.4\ \mathrm{GeV}$, as
well as estimates for $\Gamma(Z \to J/\psi \pi)+ \Gamma(Z \to \psi^{\prime}
\pi)=147.5\ \mathrm{MeV}$ and for the ratio $R_Z=4.48$ are close to our
results.

We have tested a suggestion on $Z(4430)$ resonance as first radially excited
state of the diquark-antidiquark state $Z_c(3900)$. For the masses and total
widths of the $Z_c(3900)$ and $Z(4430)$ resonances we have found: $%
m_{Z_c}=3901^{+125}_{-148}\ \mathrm{MeV}$, $\Gamma_{Z_c}=(67.3 \pm 10.8)\
\mathrm{MeV}$, and $m_{Z}=4452^{+182}_{-228}\ \mathrm{MeV}$, $%
\Gamma_{Z}=(169.7 \pm 39.2)\ \mathrm{MeV}$. Results of the present work seem
support assumption on excited nature of the $Z(4430)$ resonance. But there
are problems to be addressed before drawing strong conclusions. Namely,
theoretical investigations of other decay channels of $Z_c(3900)$ and $%
Z(4430)$ states has to be carried out in order to obtain more accurate
predictions for their total widths. Experimental studies of the $Z(4430)$
resonance's decay channels, especially a direct measurement of $\Gamma(Z \to
J/\psi \pi)$ may help in clarifying its nature as a radial excitation of $%
Z_c(3900)$ state.

\section*{ACKNOWLEDGEMENTS}

K.~A.~ thanks T\"{U}BITAK for the partial financial support provided under
Grant No. 115F183.

%%%%%%%%%%%%%%%%%%%%%%%%%%%%%%%%%%%%%%%%%%%%%%%%%%%%%%%%%%%%%%
%Appendix
%%%%%%%%%%%%%%%%%%%%%%%%%%%%%%%%%%%%%%%%%%%%%%%%%%%%%%%%%%%%%%
\appendix*

\section{ The quark propagators and spectral densities}

\renewcommand{\theequation}{\Alph{section}.\arabic{equation}} \label{sec:App}

The light and heavy quark propagators are necessary to find QCD side of the
correlation functions in both mass, current and strong couplings'
calculations. In the present work we employ the $q$- quark propagator $%
S_{q}^{ab}(x)$, which is given by the following formula%
\begin{eqnarray}
&&S_{q}^{ab}(x)=i\delta _{ab}\frac{\slashed x}{2\pi ^{2}x^{4}}-\delta _{ab}%
\frac{m_{q}}{4\pi ^{2}x^{2}}-\delta _{ab}\frac{\langle \overline{q}q\rangle
}{12}  \notag \\
&&+i\delta _{ab}\frac{\slashed xm_{q}\langle \overline{q}q\rangle }{48}%
-\delta _{ab}\frac{x^{2}}{192}\langle \overline{q}g_{s}\sigma Gq\rangle
+i\delta _{ab}\frac{x^{2}\slashed xm_{q}}{1152}  \notag \\
&&\times \langle \overline{q}g_{s}\sigma Gq\rangle -i\frac{%
g_{s}G_{ab}^{\alpha \beta }}{32\pi ^{2}x^{2}}\left[ \slashed x{\sigma
_{\alpha \beta }+\sigma _{\alpha \beta }}\slashed x\right]  \notag \\
&&-i\delta _{ab}\frac{x^{2}\slashed xg_{s}^{2}\langle \overline{q}q\rangle
^{2}}{7776}-\delta _{ab}\frac{x^{4}\langle \overline{q}q\rangle \langle
g_{s}^{2}G^{2}\rangle }{27648}+\ldots  \label{eq:qprop}
\end{eqnarray}%
For the $c$-quark propagator $S_{c}^{ab}(x)$ we use the well-known
expression
\begin{eqnarray}
&&S_{c}^{ab}(x)=i\int \frac{d^{4}k}{(2\pi )^{4}}e^{-ikx}\Bigg \{\frac{\delta
_{ab}\left( {\slashed k}+m_{c}\right) }{k^{2}-m_{c}^{2}}  \notag \\
&&-\frac{g_{s}G_{ab}^{\alpha \beta }}{4}\frac{\sigma _{\alpha \beta }\left( {%
\slashed k}+m_{c}\right) +\left( {\slashed k}+m_{c}\right) \sigma _{\alpha
\beta }}{(k^{2}-m_{c}^{2})^{2}}  \notag \\
&&+\frac{g_{s}^{2}G^{2}}{12}\delta _{ab}m_{c}\frac{k^{2}+m_{c}{\slashed k}}{%
(k^{2}-m_{c}^{2})^{4}}+\frac{g_{s}^{3}G^{3}}{48}\delta _{ab}\frac{\left( {%
\slashed k}+m_{c}\right) }{(k^{2}-m_{c}^{2})^{6}}  \notag \\
&&\times \left[ {\slashed k}\left( k^{2}-3m_{c}^{2}\right) +2m_{c}\left(
2k^{2}-m_{c}^{2}\right) \right] \left( {\slashed k}+m_{c}\right) +\ldots %
\Bigg \}.  \notag \\
&&{}  \label{eq:Qprop}
\end{eqnarray}%
In Eqs.\ (\ref{eq:qprop}) and (\ref{eq:Qprop}) we adopt the notations
\begin{eqnarray}
&&G_{ab}^{\alpha \beta }=G_{A}^{\alpha \beta
}t_{ab}^{A},\,\,~~G^{2}=G_{\alpha \beta }^{A}G_{\alpha \beta }^{A},  \notag
\\
&&G^{3}=\,\,f^{ABC}G_{\mu \nu }^{A}G_{\nu \delta }^{B}G_{\delta \mu }^{C},
\label{eq:A.3}
\end{eqnarray}%
with $a,\,b=1,2,3$ being the color indices, and $A,B,C=1,\,2\,\ldots 8$ \ .
In Eq.\ (\ref{eq:A.3}) $t^{A}=\lambda ^{A}/2$, $\lambda ^{A}$ are the
Gell-Mann matrices, and the gluon field strength tensor $G_{\alpha \beta
}^{A}\equiv G_{\alpha \beta }^{A}(0)$ is fixed at $x=0$.

The nonperturbative part of the spectral density Eq.\ (\ref{eq:SD1}) is
determined by the formula
\begin{eqnarray}
&&\rho ^{\mathrm{n.-pert.}}(s)=\Big \langle\frac{\alpha _{s}G^{2}}{\pi }\Big
\rangle m_{c}^{2}\int_{0}^{1}f_{1}(z,s)dz  \notag \\
&&+\Big \langle g_{s}^{3}G^{3}\Big \rangle\int_{0}^{1}f_{2}(z,s)dz  \notag \\
&&+\Big \langle\frac{\alpha _{s}G^{2}}{\pi }\Big \rangle^{2}m_{c}^{2}%
\int_{0}^{1}f_{3}(z,s)dz.  \label{eq:NPert}
\end{eqnarray}%
In Eq.\ (\ref{eq:NPert}) the functions $f_{1}(z,s),\ f_{2}(z,s)$ and $%
f_{3}(z,s)$ have the explicit forms:
\begin{eqnarray}
&&f_{1}(z,s)=\frac{1}{6r^{2}}\left\{ -\left( 1+3r\right) \delta
^{\prime}(s-\Phi )\right.  \notag \\
&&\left. +s(1+2r)\delta ^{(2)}(s-\Phi )\right\} ,
\end{eqnarray}%
\begin{eqnarray}
&&f_{2}(z,s)=\frac{1}{15\cdot 2^{7}\pi ^{2}r^{5}}\left\{ 2r^{2}\left[
-9m_{c}^{2}\left( 1+5r(1+r)\right) \right. \right.  \notag \\
&&\left. +2sr(3+r(19+27r))\right] \delta ^{(2)}(s-\Phi )+r\left[
s^{2}r^{3}(8+27r)\right.  \notag \\
&&\left. m_{c}^{4}(1+5r(1+r))+6m_{c}^{2}sr(3+r(11+3r))\right] \delta
^{(3)}(s-\Phi )  \notag \\
&&+\left[ s^{3}r^{5}+6m_{c}^{2}s^{2}r^{3}(1+2r)+2m_{c}^{6}(1+5r(1+r))\right.
\notag \\
&&\left. \left. -m_{c}^{4}sr(7+r(31+23r))\right] \delta ^{(4)}(s-\Phi
)\right\} ,
\end{eqnarray}%
and
\begin{eqnarray}
&&f_{3}(z,s) =\frac{\pi ^{2}}{108r^{2}}\left[ 6r\delta ^{(3)}(s-\Phi
)-(m_{c}^{2}-2s-6sr)\right.  \notag \\
&&\left. \times \delta ^{(4)}(s-\Phi )-s(m_{c}^{2}-rs)\delta ^{(5)}(s-\Phi )
\right] ,
\end{eqnarray}%
where we use the notations%
\begin{equation}
\ r=z(z-1),\ \ \Phi =\frac{m_{c}^{2}}{z(1-z)}.
\end{equation}%
In the expressions above the Dirac delta function $\ \delta ^{(n)}(s-\Phi )$
is defined as
\begin{equation}
\delta ^{(n)}(s-\Phi )=\frac{d^{n}}{ds^{n}}\delta (s-\Phi ).
\end{equation}

The nonperturbative component of $\rho^{\mathrm{QCD}}(s)$ defined by Eq.\ (%
\ref{eq:SD2}) is given by the formulas:
\begin{eqnarray}
&&\rho ^{\mathrm{n.-pert.}}(s) =\frac{\zeta _{4\rho }m_{\rho }^{2}}{s}+\Big
\langle\frac{\alpha _{s}G^{2}}{\pi }\Big \rangle m_{c}^{2}\int_{0}^{1}%
\widetilde{f}_{1}(z,s)dz  \notag \\
&&+\Big \langle g_{s}^{3}G^{3}\Big \rangle\int_{0}^{1}\widetilde{f}%
_{2}(z,s)dz  \notag \\
&&+\Big \langle\frac{\alpha _{s}G^{2}}{\pi }\Big \rangle^{2}m_{c}^{2}%
\int_{0}^{1}\widetilde{f}_{3}(z,s)dz,
\end{eqnarray}%
where
\begin{equation}
\widetilde{f}_{1}(z,s)=-\frac{s(1+2r)}{9}\delta ^{(2)}(s-\Phi ),
\end{equation}%
\begin{eqnarray}
&&\widetilde{f}_{2}(z,s) =\frac{1}{45\cdot 2^{6}\pi ^{2}r^{5}}\left\{
12sr^{3}\left[ 1+r(7+11r)\right] \right.  \notag \\
&&\times \delta ^{(2)}(s-\Phi )+2r\left[ 2s^{2}r^{3}(2+7r)-3m_{c}^{4}\left(
1\right. \right.  \notag \\
&&\left. \left. +5r(1+r)\right) +9m_{c}^{2}sr\left( 1+2r(2+z)\right) \right]
\delta ^{(3)}(s-\Phi )  \notag \\
&&+\left[ s^{3}r^{5}+6m_{c}^{2}s^{2}r^{3}(1+2r)+2m_{c}^{6}\left(
1+5r(1+r)\right) \right.  \notag \\
&&\left. \left. -m_{c}^{4}sr\left( 7+r(31+23r)\right) \right] \delta
^{(4)}(s-\Phi )\right\} ,
\end{eqnarray}%
and
\begin{eqnarray}
&&\widetilde{f}_{3}(z,s) =\frac{1}{3^{4}\cdot 2}\left\{ -6r\delta
^{(3)}(s-\Phi )+2\left[ m_{c}^{2}-s(1+3r)\right] \right.  \notag \\
&&\left. \times\delta ^{(4)}(s-\Phi)+s(m_{c}^{2}-sr)\delta ^{(5)}(s-\Phi
)\right\} .
\end{eqnarray}

\end{document}